\documentclass[twocolumn,amsmath,amssymb,prl,aps,superscriptaddress]{revtex4-2}
\usepackage[version=3]{mhchem} % Formula subscripts using \ce{}
\usepackage{amsmath}
\usepackage{upgreek}
\usepackage{color}
\usepackage{textgreek}
\usepackage{graphicx,subfigure}% Include figure files
\usepackage{hyperref}
\usepackage{nicefrac}
\usepackage{dcolumn}% Align table columns on decimal point
\usepackage{bm}% bold math
\usepackage[normalem]{ulem}
\usepackage{xcolor}

\usepackage{xspace}

\begin{document}

\title{Light-regulated adsorption and desorption of \textit{Chlamydomonas} cells at surfaces}

\author{Rodrigo E.\ Catalan}
\affiliation{Max Planck Institute for Dynamics and Self-Organization (MPIDS), Am Fa{\ss}berg 17, D-37077 G\"{o}ttingen, Germany}
\affiliation{Experimental Physics V, University of Bayreuth, Universit\"atsstr.\ 30, 95447 Bayreuth, Germany}

\author{Alexandros A.\ Fragkopoulos}
\affiliation{Max Planck Institute for Dynamics and Self-Organization (MPIDS), Am Fa{\ss}berg 17, D-37077 G\"{o}ttingen, Germany}
\affiliation{Experimental Physics V, University of Bayreuth, Universit\"atsstr.\ 30, 95447 Bayreuth, Germany}

\author{Nicolas von Trott}
\affiliation{Max Planck Institute for Dynamics and Self-Organization (MPIDS), Am Fa{\ss}berg 17, D-37077 G\"{o}ttingen, Germany}

\author{Simon Kelterborn}
\affiliation{Institute of Biology, Experimental Biophysics, Humboldt-Universit\"{a}t, Invalidenstraße~42, 10115 Berlin, Germany}
\affiliation{Institute of Translational Physiology, Charité - Universit\"{a}tsmedizin Berlin, corporate member of Freie Universit\"{a}t Berlin and Humboldt-Universit\"{a}t zu Berlin, 10117 Berlin, Germany}

\author{Olga Baidukova}
\affiliation{Institute of Biology, Experimental Biophysics, Humboldt-Universit\"{a}t, Invalidenstraße~42, 10115 Berlin, Germany}

\author{Peter Hegemann}
\affiliation{Institute of Biology, Experimental Biophysics, Humboldt-Universit\"{a}t, Invalidenstraße~42, 10115 Berlin, Germany}

\author{Oliver B\"{a}umchen}
\thanks{E-mail: oliver.baeumchen@uni-bayreuth.de}
\affiliation{Max Planck Institute for Dynamics and Self-Organization (MPIDS), Am Fa{\ss}berg 17, D-37077 G\"{o}ttingen, Germany}
\affiliation{Experimental Physics V, University of Bayreuth, Universit\"atsstr.\ 30, D-95447 Bayreuth, Germany}

\begin{abstract} 
\noindent Microbial colonization of surfaces represents the first step towards biofilm formation, which is a recurring phenomenon in nature with beneficial and detrimental implications in technological and medical settings. 
Consequently, there is a current interest in elucidating the fundamental aspects of the initial stages of biofilm formation of microorganisms on solid surfaces. 
While most of the research is oriented to understand bacterial surface colonization, such observations at a fundamental level using photosynthetic microalgae are thus far elusive. 
Recent single-cell studies showed that the flagellar adhesion of \textit{Chlamydomonas} is switched on in blue light and switched off under red light\ [Kreis \textit{et al., Nature Physics}, 2018, \textbf{14}, 45-49]. 
Here, we study this light-switchable surface association of \textit{C.~reinhardtii} on the population level and measure the kinetics of adsorption and desorption of suspensions of motile cells on glass surfaces using bright field optical microscopy. 
We observe that both processes exhibit a response lag relative to the time at which the blue- and red-light conditions are set and model this feature using time-delayed Langmuir-type kinetics. 
We find that cell adsorption occurs significantly faster than desorption, which we attribute to the protein-mediated molecular adhesion mechanism of the cells. 
Adsorption experiments using phototactically blind \textit{Chlamydomonas} mutants demonstrate that phototaxis does not affect the cell adsorption kinetics. 
Hence, this method can be used as an assay for characterizing the dynamics of the surface colonization of microbial species exhibiting light-regulated surface adhesion.

\end{abstract}

\maketitle

%%%MAIN TEXT%%%%
%The main text of the article\cite{Mena2000} should appear here.

The development of microbial colonies at natural and artificial surfaces, known as biofilms, is a recurring phenomenon that has already been found in ancient forms of life, such as in microbial mats\cite{noffke2013microbially}, microfossils\cite{rasmussen2000filamentous}, and even more presently in medical settings such as dental caries\cite{li2004identification}, mucosal infections\cite{motta2021gastrointestinal} and bacterial contamination of artificial implants\cite{stickler2008bacterial}. 
Biofilms are vital communities of microorganism, in which the cells are protected by a self-produced matrix of extracellular polymeric substances (EPS)\cite{flemming2016biofilms}, and can adhere to one another as well as to solid surfaces\cite{o2000biofilm}. 
These populations of cells can host multiple different species \cite{flemming2016biofilms} and are considered as a dynamic and complex biological system with emergent properties that provide essential survival advantages to its community \cite{stewart2001antibiotic,mah2001mechanisms,bridier2011resistance,hall2004bacterial}. 
As a result of these collective properties, the formation of microbial colonies at surfaces has far-reaching implications in economical, technological, and medical settings\cite{costerton1999,nicolella2000,van1993,kumar1998}. 
The interest in both preventing\cite{francolini2010prevention} and promoting\cite{thompson2002importance} the formation of biofilms in technological applications and physiological environments has stimulated numerous studies aiming at elucidating the conditions and mechanisms by which cells interact, settle and detach from surfaces. 
Particularly important for such applications are the initial stages of biofilm formation, which involve approach, surface sensing\cite{armbruster2018,utada2014vibrio} and attachment to surfaces\cite{marshall1971mechanism}. 
Despite the tremendous advancement achieved in recognizing the mechanisms involved in the formation of biofilms, the vast majority of the literature centers around bacterial surface colonization\cite{loskill2012,maikranz2020different,dufrene2020mechanomicrobiology,marshall2006,diaz2007,sun2020,lee2018,mazza2016}. 
In contrast, studies of biofilms in other important exemplars of microbial life, particularly in photoactive microalgae, remain rather elusive.
Microalgae are a diverse group of eukaryotic and photosynthetic organisms that are known to be primary producers of oxygen and organic molecules on Earth and, thus, are a fundamental support for the existing ecosystems\cite{beardall2002ecological}. 
These microorganisms can be found in their planktonic, i.e.\ free-swimming, state in freshwater and marine ecosystems, but also colonizing natural and artificial light-exposed moist surfaces. 
Particularly for artificial surfaces, microalgae can also have non-desired implications if a biofilm community is established\cite{lamenti2000biodeterioration,schultz2011economic}. 
Regarding these issues, microalgae have attracted the interest of researchers for their application in wastewater treatment \cite{boelee2011,berner2015} as well as in photobioreactors for the production of biofuels and synthesis of pharmaceutical components \cite{zeriouh2017biofouling,christenson2011production}. 
In contrast to detailed studies of bacterial surface colonization, there have been mostly qualitative studies of microalgae focusing on specific applications directly, thus leaving fundamental aspects rather unconsidered\cite{roostaei2018mixotrophic,schumann2005chlorophyll}.
The unicellular soil-dwelling microalga \textit{Chlamydomonas reinhardtii} is a model organism\cite{harris2001chlamydomonas} that has been extensively employed to study fundamental biological and biophysical processes, such as photosynthesis\cite{hippler1998chlamydomonas,rochaix2006molecular}, flagellar assembly\cite{salome2019series} and coordination\cite{goldstein2011,geyer2013,wan2016,boeddeker2020} as well as microbial motility\cite{polin2009,ostapenko2018,cammann2021}. 
Its flagella do not only play a fundamental role in mating, but also allow individuals to swim through their fluid environment and to interact with surfaces.
Surface association is enabled by adhesive contacts between the surface and glycoproteins, known as FMG-1B, that are localized along the flagellar membrane\cite{bloodgood1984flagellar}. 
This glycoprotein, in conjunction with the transport of motor proteins along the flagellar axoneme, enables the cells to glide along the surface\cite{bloodgood2019chlamydomonas}.
In their surface-associated state, the flagella are non-motile, typically wide-spread and oriented at about 180$^\circ$ to one another, known as gliding configuration. 
The gliding motility itself is then bidirectional along the flagella\cite{bloodgood1981gliding}.
Interestingly, the flagellar adhesion of individual \textit{C.~reinhardtii} cells to surfaces can be switched off under red-light conditions, while it fully recovers under blue light\cite{kreis2018adhesion,kreis2019vivo,xu2020}. 

In this work, we exploit the light-switchable flagellar adhesion of \textit{C.~reinhardtii} in a cell suspension and characterize the kinetics governing the early stage of surface colonization. 
For this, we use optical microscopy and cell detection and tracking to monitor the temporal dependence of the number of cells adhered to a glass substrate. 
We propose an extended Langmuir adsorption model, which includes the average time it takes for cells to change their adhesive properties, that captures the experimental data.
We also quantify the relevant temporal parameters governing the surface colonization of the cells for different light intensities above the critical light intensity threshold for surface adhesion.
In light of the fact that \textit{C.~reinhardtii} is also able to sense light gradients and freely swim towards or away from a source of light, i.e\ perform phototaxis\cite{stavis1973phototaxis}, we assess the effect of this phototactic response on our measurements by means of specific photoreceptor deletion mutants. Finally, we show that when phototaxis is inhibited, the natural swimming of the cells against the gravity gradient (i.e. negative gravitaxis\cite{bean1977geotactic}) dictates the boundary of the experimental compartment at which most of the cells in the suspension will adsorb.\\

%\subsection{This is the subsection heading style}
%Section headings can be typeset with and without numbers.\cite{Abernethy2003}

%\subsubsection{This is the subsubsection style.~~} These headings should end in a full point.  

%\paragraph{This is the next level heading.~~} For this level please use \texttt{\textbackslash paragraph}. These headings should also end in a full point.

\section{Material and Methods}
\subsection{Cell cultivation}
Wild-type (WT) \textit{C.~reinhardtii} cells, strain SAG11-32b, and channelrhodopsin-1 and -2 double knockouts (\textDelta ChR1,2) of the WT strain were grown axenically in Tris-acetate-phosphate (TAP) medium (Gibco\textregistered, Waltham, Massachusetts, USA) on a 12 h/12 h day-night cycle in the controlled temperature and light conditions of a Memmert IPP 110Plus incubator. 
The daytime temperature was 24\,$^\circ$C with light intensity (white LED) of $1\cdot10^{20}-2\cdot10^{20}$~photons$\cdot$m$^{-2}\cdot$s$^{-1}$. 
The temperature during the dark cycle was 22ºC with the light intensity reduced to zero.
Further information regarding the \textDelta ChR1,2 strain as well as the WT strain are now available under the label CC\nobreakdash-5679 at the Chlamydomonas Resource Center (\url{https://www.chlamycollection.org/}).

\subsection{Chamber production}
Circular compartments of polydimethylsiloxane (PDMS; Sylgard\texttrademark 184, Dow Corning, Michigan, USA) were made by mixing the base and curing agent with a 10:1 weight ratio respectively, as recommended by the manufacturer.
After mixing, the product was degassed in a vacuum chamber. 
The mixture was placed between two glass slides separated by a stack of three spacers, each of 100\,\textmu m of height. 
The glass slides were placed in an oven at 75\,$^\circ$C for 2 hours. 
After curing, the PDMS slabs were removed and their height was measured to be 300$\pm$20\,\textmu m. 
A Harris uni-core punch was used to cut holes of 4~mm in diameter in the PDMS slabs.

\subsection{Sample preparation}

\begin{figure}[ht]
    \centering
    \includegraphics{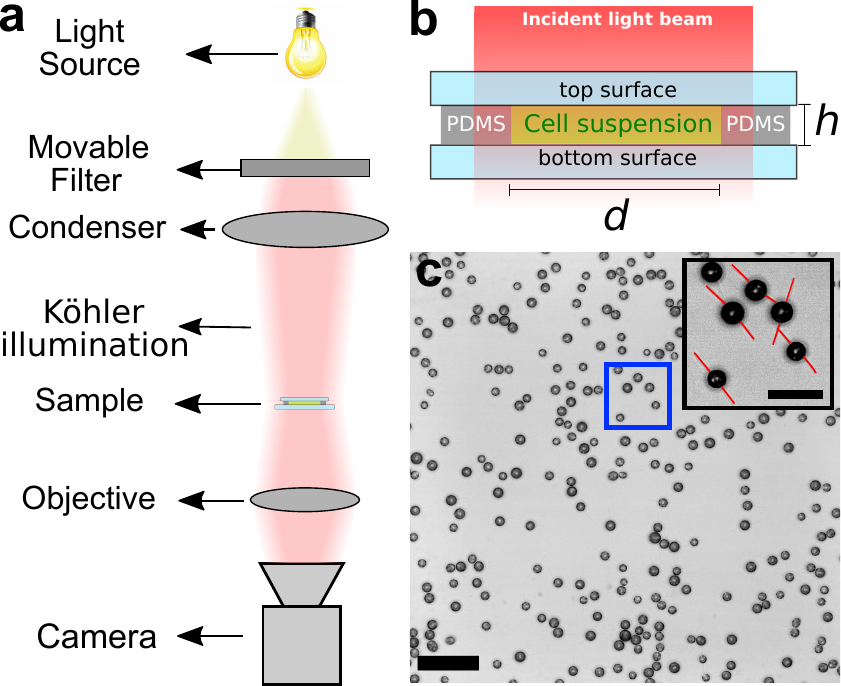}
    \caption{(a) Schematics of the experimental setup. (b) Side view of the circular PDMS compartment containing the cell suspension. The inner diameter of the chamber is $d=4$~mm and the height is $h=300$~\textmu m. (c) Optical micrograph obtained 240~s after switching to blue light showing \textit{C.~reinhardtii} cells adsorbed at the bottom surface of the compartment during a representative adsorption experiment (scale bar is 50~\textmu m, magnification 32x). The inset micrograph is a close-up of the cells attached at the area enclosed in blue. The flagella have been highlighted in red for guidance (scale bar is 20~\textmu m).}.
    \label{fig:setup}
\end{figure}

%Cell centrifugation and controlled cell density
For the adsorption experiments, vegetative \textit{C.~reinhardtii} cells were taken from the cultures in their logarithmic growth phase during mid daytime on the third day after incubation. 
In order to work with a controlled cell density, we counted the number of cells in a small volume of cell suspension. 
For this, 50~mL of each culture were centrifuged at 100~g for ten minutes (centrifuge 5804R, Eppendorf, Hamburg, Germany). 
After centrifugation, around 49~mL of the excess liquid phase was removed and 1~mL of fresh TAP was added to resuspend the cells. 
Then the cultures were placed in an incubator at 24\,$^\circ$C for 1-2 hours to ensure full regrowth of the flagella\cite{randall1968developmental,rosenbaum1969flagellar}.\\
%Cell counting and sample chamber
A volume of 1~mL was taken from the top of the resuspended culture to be used in the experiments in order to remove as much as possible the presence of dead cells. 
Finally, the cell density was determined by using a hemocytometer (Neubauer Improved; Laboroptik Ltd, Lancing, UK). 
All the experimental suspensions were then diluted to obtain a controlled density of $\left(5.0\pm0.4\right)\cdot10^{6}$~cells/mL. \\
A PDMS chamber was placed on a clean glass slide and filled with a volume of 80~$\mu$L of this suspension. 
In order to avoid evaporation in the suspension during experiments, the cell chamber was closed by placing another glass slide on top of the PDMS compartment (see Fig \ref{fig:setup}b). 
The suspension was stirred before to remove density inhomogeneities due to the cell's natural gravitaxis\cite{kam1999gravitaxis} and phototaxis\cite{arrieta2017phototaxis}. 
Any excess of liquid was removed after closing the chamber.

\graphicspath{{Figures/}}

\graphicspath{{Figures/}}
\begin{figure*}[t]
    \centering
    \includegraphics[width=\linewidth]{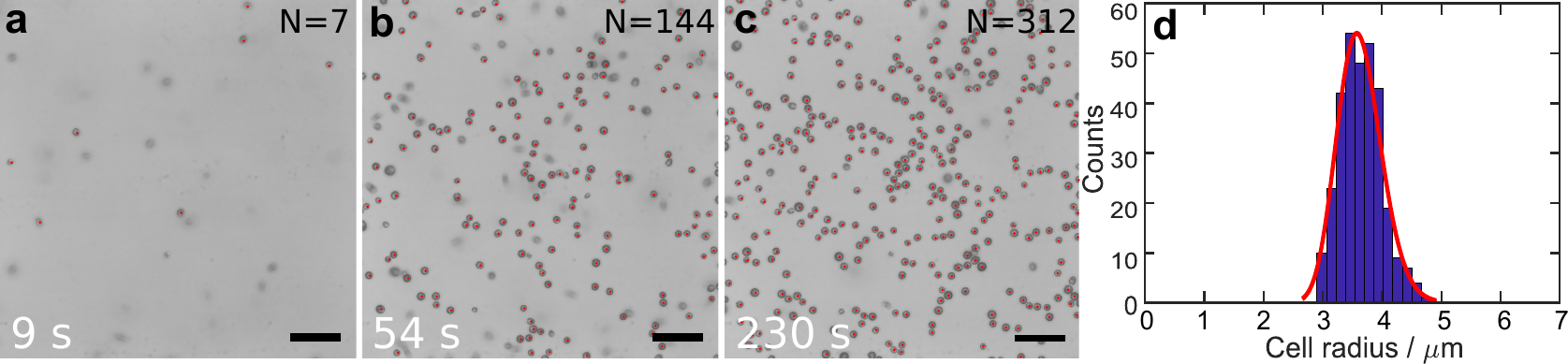}
    \caption{(a-c) Series of optical micrographs showing adsorbed WT \textit{C.~reinhardtii} cells during a representative adsorption run. Time stamps indicate the time after switching to blue light conditions. Successfully detected cells are marked with a red dot. The scale bar is 50~\textmu m. (d) Histogram depicting the radius of adsorbed cells (N=312, 12 bins) as obtained from the cell tracking algorithm. The solid (red) line represents a best fit to a log-normal distribution for an average particle radius of 3.61$\pm$0.06~\textmu m.}
    \label{fig:detection}
\end{figure*}

\subsection{Adsorption experiments}
%Light illumination and incubation
The chamber containing the cell suspension was placed on the stage of an inverted optical microscope (IX-83, Olympus Corporation, Japan) and observed under bright-field microscopy, see Fig.\ref{fig:setup}a. 
The cell suspension was incubated in red light for 15 minutes using a bandpass interference filter ($\lambda$=664~nm, FWHM = 11~nm). 
The illumination was provided by a LED system with an photon flux of $1\cdot10^{19}$~photons$\cdot$m$^{-2}\cdot$s$^{-1}$. 
The dimensions of the cell chamber are such that, after setting up K\"{o}hler illumination, the light intensity of the collimated beams is homogeneous within the chamber.

After incubation, we replaced the red filter by a blue bandpass interference filter ($\lambda$= 476~nm, FWHM = 11~nm) and recorded time series of bright-field micrographs of the adsorption of \textit{C.~reinhardtii} at 3~fps, focusing on the bottom surface at the center of the compartment (see Fig.\ref{fig:setup}c). 
Images were recorded for 330~s under blue light before switching back to red illumination, during which the cells were recorded for another 330~s to study the desorption of the cells from the bottom surface. 
The time span between one cycle of measurements and the next was about 10 minutes. 
The light intensity between the two light conditions was kept constant with a corresponding photon flux of 1$\cdot$10$^{19}$~photons$\cdot$m$^{-2}\cdot$s$^{-1}$ in each cycle. 
In order to study the effect of the light intensity on the adsorption kinetics, we performed experiments for which, in each cycle, the cells were exposed to one of four light intensities (0.5, 0.8, 1, and 2$\cdot$10$^{19}$~photons$\cdot$m$^{-2}\cdot$s$^{-1}$. 
The intensities were randomly assigned in order to avoid adaptation effects. 
The cells were exposed to the desired light intensity for 10 minutes in red light before the start of each adsorption-desorption cycle. 
At the end of each cycle, the cells were exposed to darkness for 1-2 minutes so that most of the cells remaining on the surface could swim back to the bulk.
To assess potential effects originating from phototaxis in our experiments, we studied the adsorption kinetics of genetically modified SAG11-32b cells, which lack the two blue-light photoreceptors\cite{greiner2017targeting}, channelrhodopsin-1 (ChR1)\cite{nagel2002channelrhodopsin} and channelrhodopsin-2 (ChR2)\cite{nagel2003channelrhodopsin}, which are known to account for phototactic responses of \textit{C.~reinhardtii}. 
We then compared the surface density of cells adsorbed on the top surface of the compartment with the density of cells adsorbed on the bottom one.

\subsection{Image analysis}
An ideal requirement to analyze a 2D-collection of particles in a set of micrographs is to have enough resolution and contrast between the suspension medium and the particles to allow for their spatial detection. 
Several methods and software have been developed to perform these tasks, each of them having their own criteria and accuracy\cite{chenouard2014objective,meijering2012methods}.
%Image thresholding is not useful in our experiments
One of the most prominent ones, especially used when particles have sufficiently different intensities than their surroundings, is the use of thresholding. 
This method consists in choosing an optimal value of pixel intensity to generate a binary image displaying disconnected regions labeled as 'particles' and a connected region labeled as 'background'. 
However, this method is not effective when the particles exhibit a non-uniform contrast respect to the background. 
In our experiments, the 32x magnification may resolve the complex structures and organelles inside the cell bodies. 
Complex structures scatter the light passing through the cell body, which makes the determination of the contrast between the cells and the background inaccurate.

%Circular Hough Transform
Since the cell bodies of \textit{C.~reinhardtii} appear circular when they are adhered on the surface with their flagella facing in opposite directions, also known as gliding configuration, we apply the \textit{imfindcircles} MATLAB algorithm, which uses the circular Hough transform to locate circular objects given an interval of pixel radii\cite{duda1972use}.
For our image analysis, we use pixel radii ranging between 15 and 40 pixels which is equivalent to radii between 2.6~\textmu m and 7~\textmu m. 
The results of the cell detection are shown in Figure \ref{fig:detection}(a-c), which displays micrographs corresponding to three specific instants of an adsorption essay, namely 9~s, 54~s and 230~s after exposure to blue light. 
The detected cells, which are adsorbed at the bottom surface of the circular compartment, are considered as circles and thus highlighted in red. 
Cells swimming in the vicinity of the surface appear blurry and are not detected by the algorithm in general. 
The circular Hough transform provides information about the radii of the detected objects. 
We find that the cell radii follow a log-normal distribution with a mean of 3.63~\textmu m and a standard deviation 0.06~\textmu m.

%Track function (E. Dufresne)
During blue-light conditions, when the cells are adsorbed on the surface, most of them stay in the gliding configuration, in which they move on the surface along the flagella direction \cite{bloodgood1981gliding}. 
However, some cells remain only loosely attached and may transit back to the planktonic cell suspension. 
Those cells, which are not completely attached to the surface, might still be detected as adsorbed cells by the algorithm, leading to noisy data or to a miscount in the number of cells in the gliding configuration. 
Thus, once the algorithm has detected the cells on the surface, it is necessary to discriminate which ones are firmly attached to it and which are not. 
To accomplish this, we monitor the position of each detected cell over subsequent micrographs using a MATLAB tracking algorithm\cite{blair2008matlab}, which links the location of all cells throughout the frames to form a trajectory. 
Since the gliding speed of cells is around 1-2~\textmu m/s\cite{shih2013intraflagellar}, the displacements between recording frames taken at 3 fps are around 0.7~\textmu m, which is less than the average cell radius shown in Fig.~\ref{fig:detection}d. 
Thus we allow the tracking function to consider displacements up to one average cell radius between each recording frame, which is enough to count moving cells once they are adsorbed to the surface.
Cells that are only tracked for less than 2~s are sorted out as well, so we end up counting only the cells firmly attached to the surface. 
By identifying the adhered cells, we monitored the time-dependent surface cell density throughout each cycle of adsorption and desorption.\\

\section{Results and discussion}
\begin{figure}[ht]
    \centering
    \includegraphics{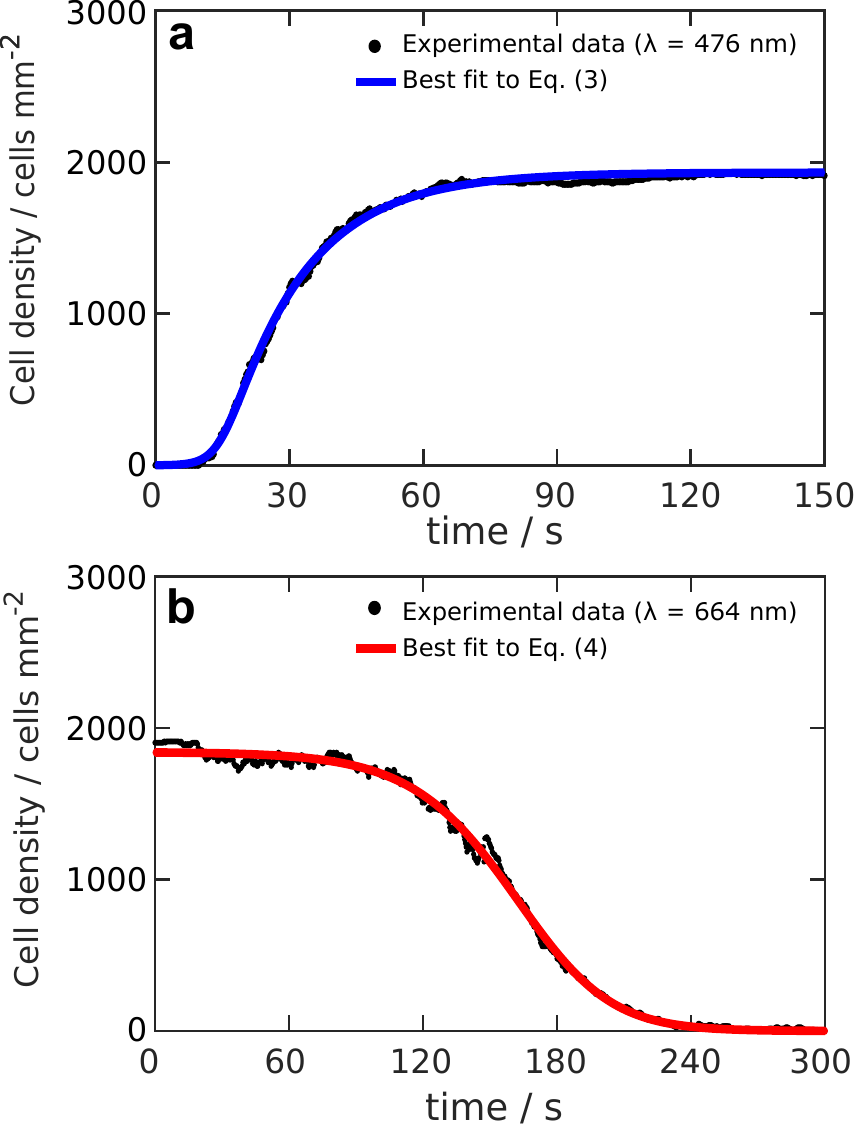}
    \caption{(a) Cell density (black dots) during adsorption and best fit of the adsorption model (blue line) given as Eq.~(\ref{eq3}). (b) Cell density (black dots) during desorption and best fit of the desorption model (red line) given as Eq.~(\ref{eq4}). For both plots, the starting time $t=0$~s represents the moment at which light conditions are changed for red to blue (adsorption) and from blue to red (desorption), respectively. The intensity of blue as well as of red light for these experiments was kept constant at $10^{19}$~photons$\cdot$m$^{-2}\cdot$s$^{-1}$}
    \label{fig:Fig3}
\end{figure}

The particle detection and tracking reveal that the dynamics of adsorption and desorption follow a monotonically increasing or decreasing function, respectively, that appears to be reminiscent of a sigmoidal curve, see Fig.\ref{fig:Fig3}.

During both adsorption and desorption, we observe that the cell surface density initially remains constant for a certain amount of time, after which it rapidly changes. 
Under blue light, this rapid change manifests as an increase of the number of cells attached to the surface until a constant plateau density is achieved. 
After switching back to red light, the change is seen as the detachment of cells from the surface, which transit back to the planktonic state. 

\subsection{Extended Langmuir-type model for adsorption and desorption}
To quantitatively capture the dynamics of both processes, we developed a model that is inspired by the Langmuir model for describing the adsorption kinetics of, e.g., molecules at surfaces\cite{langmuir1918adsorption, liu2008langmuir}. 
This model assumes that the rate of adsorption decreases as the adsorption sites on a surface are successively populated by immobile objects. 
A key difference to our system is that, \textit{C.~reinhardtii} cells can glide on the surface and potentially increase the available area for new cells to adhere.
To diminish this effect, we work with small cell densities, such that the number of adsorption sites is much larger than the number of particles, and thus the adsorption rate is limited predominantly by the number of available cells in the suspension. 
However, a caveat of the Langmuir model is that it fails to capture the characteristic initial time delay observed in our data. 
Hence, we extended the Langmuir model towards a time-dependent factor $s\left(t\right)$, such that the governing equations read:

\begin{equation}\label{eq1}
    \frac{d\sigma\left(t\right)}{dt}=\frac{1}{\tau_{_{a}}}s\left(t\right)\left(\sigma_{_{0}}-\sigma\left(t\right)\right)
\end{equation}
\begin{equation}\label{eq2}
    s\left(t\right)=\frac{1}{2}+\frac{1}{2}\tanh\left(\frac{t-\tau_{_\mathrm{delay}}}{\tau_{_{b}}}\right)
\end{equation}

In Eq.~(\ref{eq1}), $\sigma_0$ and $\sigma\left(t\right)$ represent the saturated and instantaneous surface cell density, respectively, while $\tau_{_{a}}$ is the characteristic time that determines the adsorption rate of the cells. 
Note that the classical Langmuir model is recovered for $s=1$.
The time-dependent prefactor $s\left(t\right)$ in Eq.~(\ref{eq2}) is assumed to be a smooth step-function from zero to one that we coin the \textit{stickiness function}.
It accounts for the fact that there is a significant time delay, as well as a cell-to-cell variability of this time delay, associated to the fact that the flagella of each cell in the suspension switch from the sticking to the non-sticking state\cite{kreis2018adhesion}.
In this expression for $s\left(t\right)$ displayed as Eq.~(\ref{eq2}), $\tau_{\mathrm{delay}}$ is the time at which 50\% of the cell population in the suspension has switched its adhesive properties. 
The parameter $\tau_{_{b}}$ measures the width of the step, providing a scale of the cell-to-cell variability for the time that it takes for the cells to switch their adhesive properties. 
A similar model for the desorption inferred by exchanging $\sigma_{_{0}}-\sigma\left(t\right)$ by $\sigma\left(t\right)$ in Eq.~(\ref{eq1}).
The analytical solution for the adsorption and desorption equations are respectively:

%Adsorption fit
\begin{equation}\label{eq3}
    \frac{\sigma_{_\mathrm{ads}}\left(t\right)}{\sigma_{_{0}}}=1-e^{^{\frac{-t}{2\tau_{a}}}}\cdot\cosh^{^{\frac{\tau_{b}}{2\tau_{a}}}}\left(\frac{\tau_{_\mathrm{delay}}}{\tau_{b}}\right)\cdot\cosh^{^{\frac{-\tau_{b}}{2\tau_{a}}}}\left(\frac{\tau_{_\mathrm{delay}}-t}{\tau_{b}}\right)
\end{equation}

%Desorption fit
\begin{equation}\label{eq4}
    \frac{\sigma_{_\mathrm{des}}\left(t\right)}{\sigma_{_{0}}}=e^{^{\frac{-t}{2\tau_{a}}}}\cdot\cosh^{^{\frac{\tau_{b}}{2\tau_{a}}}}\left(\frac{\tau_{_\mathrm{delay}}}{\tau_{b}}\right)\cdot\cosh^{^{\frac{-\tau_{b}}{2\tau_{a}}}}\left(\frac{\tau_{_\mathrm{delay}}-t}{\tau_{b}}\right)
\end{equation}

These analytical solutions were fitted to the experimental data in MATLAB using robust regression that minimizes the sum of the square of residuals to obtain best-fit adsorption and desorption plots, see Fig.~\ref{fig:Fig3}. 
We find that our extended Langmuir model captures the time delay observed in our experiments, allowing for a complete quantification of the dynamics. 

%In some rare cases we observed the occurrence of secondary adsorption process at longer times, in which a group of cells adsorbs to the surface at a later stage, after a primary plateau was established.
%We hypothesize that secondary adsorption is possible as adsorbed cells glide over the surface and, in consequence, expose broader areas (i.e. adsorption sites) to which new cells can attach.
%Cells contributing to secondary adsorption might have adsorbed to the upper surface of the compartment initially and swum down to the bottom surface lately. 
%The differences between primary and secondary plateau densities were not consistent between independent experiments. 
%Therefore we applied our model to study just the kinetics of the cells in the population that established the primary plateau.

\subsection{Time parameters}
\begin{figure}[ht]
    \centering
    \includegraphics{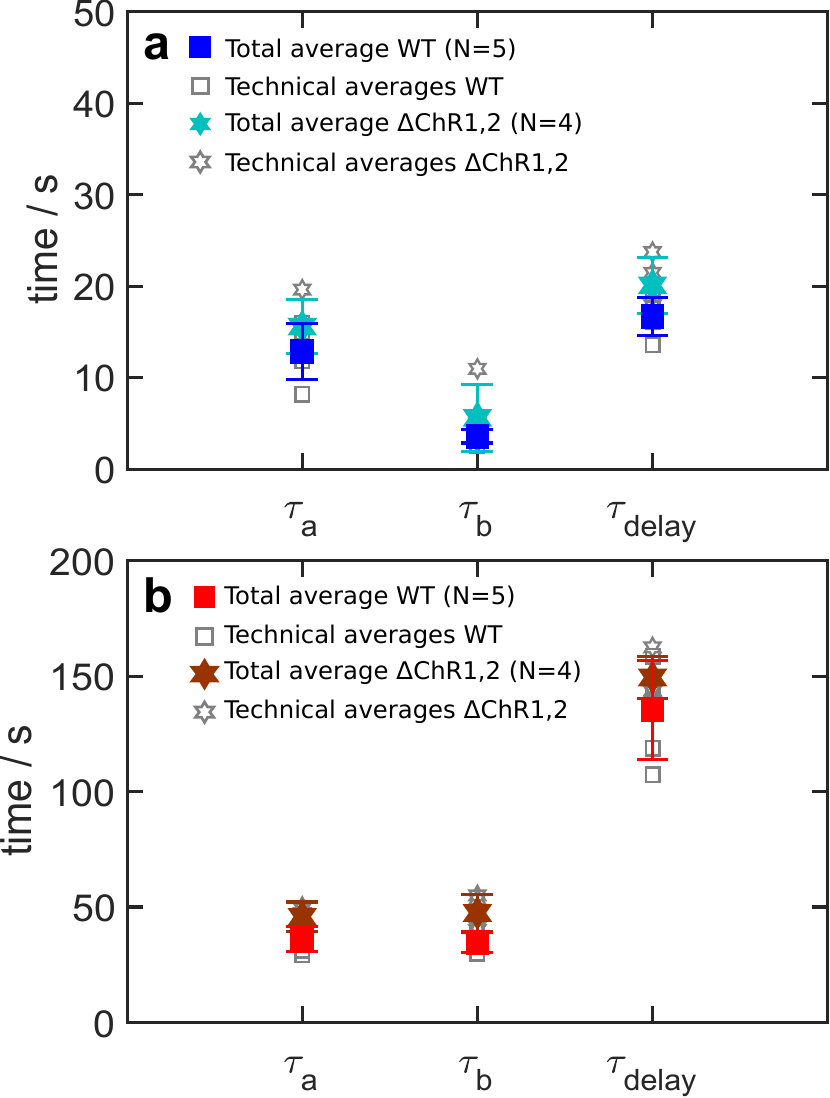}
    \caption{Comparison of the characteristic time parameters for (a) adsorption and (b) desorption of the wild-type (WT, rectangles) and channelrhodopsin-1,2 knockout (stars) strains. Measurements were taken at the bottom surface of the experimental chamber. Open symbols indicate the average of a single experiment consisting of 5 to 6 consecutive adsorption-desorption cycles. Filled symbols indicate the total average of the $N$ independent experiments performed with a light intensity of 1$\cdot$10$^{19}$~photons$\cdot$m$^{-2}\cdot$s$^{-1}$.}
    \label{fig:Fig4}
\end{figure}

The dynamics of the adsorption and desorption are notably different, as evidenced by the fact that the adsorption occurs considerably faster than the desorption, see Fig.~\ref{fig:Fig4}. 
In fact, from best fits of the extended Langmuir model we observe that the value of $\tau_{_{a}}$ for the desorption ($\tau_{_{a}}=36.0\pm5.5$~s) is, on average, about three times larger than the one for the adsorption ($\tau_{_{a}}=12.9\pm3.0$~s). 
Also the delay time are about one order of magnitude larger for the desorption ($\tau_{_\mathrm{delay}}=135.4\pm21.4$~s) as compared to the adsorption ($\tau_{_\mathrm{delay}}=16.7\pm2.1$~s). 
The difference of $\tau_{_{b}}$ between adsorption and desorption arises from its relation to the delay time. 
However, the relative variability $\tau_{_\mathrm{delay}}/\tau_{_{b}}\approx4.6$ is similar in both processes. 

%Explain the possible origin of the difference
We hypothesize that the difference in the time delay in adsorption and desorption processes could be explained considering the number of flagellar adhesion sites at which sticky proteins interact with the substrate. 
The flagellar membrane glycoprotein FMG-1B mediates the flagellar adhesion of \textit{C.~reinhardtii} to surfaces \cite{bloodgood1984flagellar}. 
For the adsorption process, the cells need only a few adhesion sites on the flagella to interact with the surface and stick to it. 
After the initial ``touch'', in which the tip of both flagella contacts the surface, comes the ``pull'' where more adhesion sites from the tip to the base of the flagella come in contact with the surface. 
The individual adsorption is complete when the total extent of each flagellum is interacting with the substrate, in the so-called gliding configuration. 
In contrary, as soon as the cells are illuminated with red light, the desorption process begins when the flagella start disabling their adhesive contacts on the surface. 
%During this process the cell bodies wiggle faster than the recording frame rate, which makes the cells appear blurry in the micrographs, hence resulting in a noisier plateau due to the loss in the number of objects detected by the algorithm, see Fig.~\ref{fig:Fig3}b. 
As the number of adhesion sites decreases the beating of the flagella recovers and becomes successively more prominent until the cells finally detach completely and swim back to the bulk suspension.

\begin{figure}[ht]
    \centering
    \includegraphics{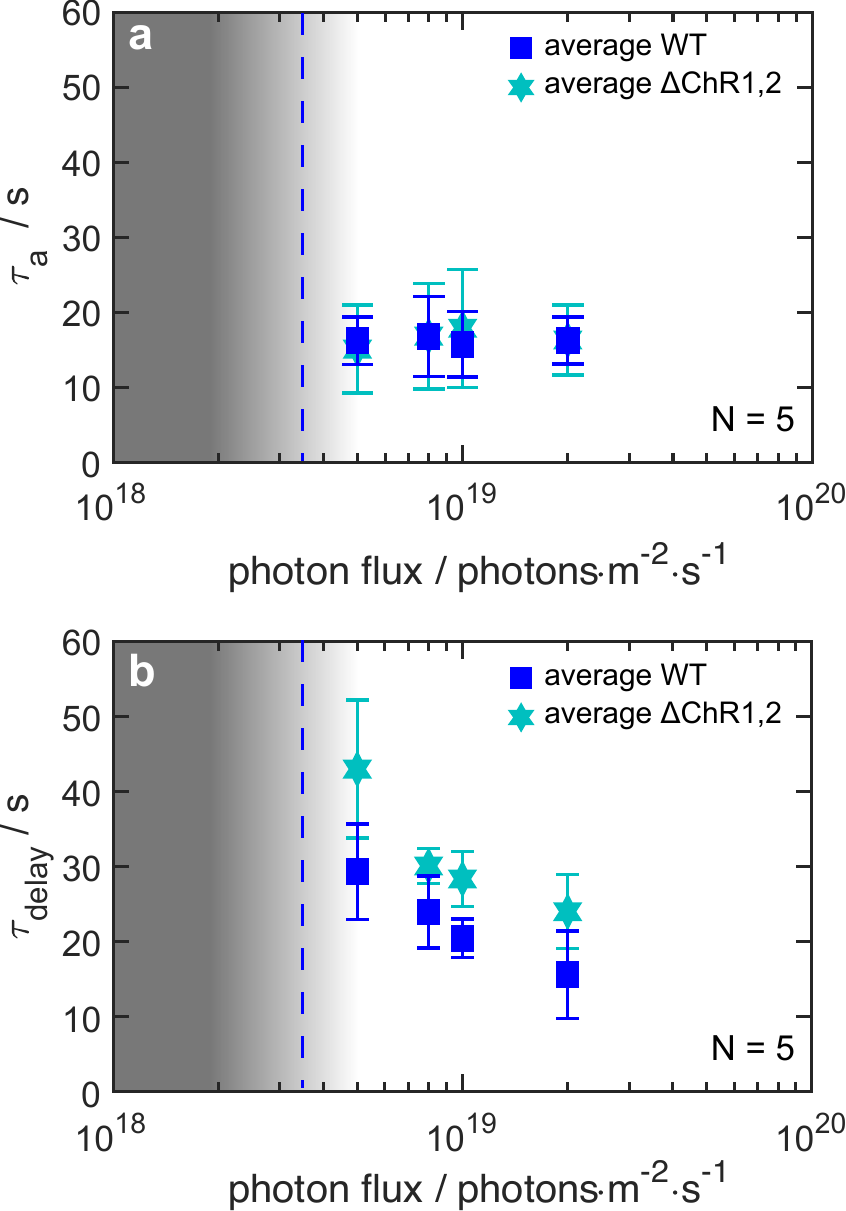}
    \caption{Light-intensity dependence of the temporal fit parameters (a) $\tau_a$ and (b) $\tau_{_\mathrm{delay}}$ for N=5 independent adsorption experiments. The gradient regions indicate the measured light intensity threshold for the light switchable adhesion\cite{kreis2018adhesion} at 2-5$\cdot10^{18}$~photons$\cdot$m$^{-2}\cdot$s$^{-1}$, with the vertical dashed blue line representing the mean. Below this range the cells do not exhibit light-induced flagellar adhesion. Measurements using the $\Delta$ChR1,2 strain were taken focusing on the top surface of the experimental compartment as most of the cells adhere at the top boundary, see Fig.\ref{fig:density_ratio}. Filled rectangles indicate average values obtained from WT experiments, whereas filled stars represent values obtained from ChR1,2-deleted mutants.}
    \label{fig:Fig5}
\end{figure}

By comparing the adsorption and desorption timescales of the channelrhodopsin-1,2 knockout strain with the ones shown by the WT, see Fig.~\ref{fig:Fig4}, we find that the time parameters $\tau_{_{a}}$ and $\tau_{_{b}}$ of the $\Delta$ChR1,2 strain are similar to the ones exhibited by WT cells.
Only $\tau_{delay}$ for the adsorption shows a systematically higher value compared to the WT (see Fig.~\ref{fig:Fig4}a, Fig.~\ref{fig:Fig5}b), which will be discussed further in the next section in the context of a light intensity sweep.

%Explain results for different intensities
\subsection{Effect of the light intensity}

Since the cells switch their adhesive properties above a well-defined intensity threshold\cite{kreis2018adhesion} between 2 and 5$\cdot10^{18}$~photons$\cdot$m$^{-2}\cdot$s$^{-1}$, we explored the light-intensity dependence of the parameters $\tau_a$ and $\tau_{_\mathrm{delay}}$ for the WT strain.
Furthermore, we independently assessed the effect of phototaxis on the adsorption parameters using ChR1,2-deleted mutants\cite{greiner2017targeting}, see Fig.~\ref{fig:Fig5}.

First, we find that the value of adsorption rate $\tau_{a}$ does not depend on the light intensity, see Fig.~\ref{fig:Fig5}a.  
%As we decrease the light intensity, it takes longer for the cells to become sticky but once they switch to the adhesive state the rate at which the population adsorbs to the surface is the same for all intensities.
The adsorption rate is mainly governed by the rate at which planktonic cells encounter the surface, which depends on the motility and the compartment's surface-to-volume ratio.
Both of these parameters are independent of the light intensity giving rise to the adsorption rate $\tau_{a}$ being independent of the light intensity.

Second, the time lag $\tau_{_\mathrm{delay}}$ monotonically decreases for increasing light intensities, see Fig.~\ref{fig:Fig5}b.
This result suggests that, in this regime, the time it takes to switch the adhesiveness of the cells on average increases as the light intensity decreases. 
Since the adhesive response of \textit{C.~reinhardtii} is light-activated, a decrease in the light intensity implies that there are less photons triggering the signalling pathway associated to switching the adhesion-state of the flagella. 
As the light intensity decreases the probability of switching to the adhesive state is lowered and, hence, the average delay time increases.
Particularly for the highest light intensity (2$\cdot10^{19}$~photons$\cdot$m$^{-2}\cdot$s$^{-1}$) used in these experiments, the time delay for the WT is found to be  15.6$\pm$5.8~s, which indeed agrees well with the typical timescale of the light-switchable flagellar adhesiveness as obtained from single-cell auto-adhesion experiments performed in white light at the same intensity\cite{kreis2018adhesion}.

Third, both the WT and the channelrhodopsin-1,2 knockout strain exhibit consistent adsorption rates above the light-intensity threshold for surface adhesion, with values of $\tau_{a}=16.4\pm0.4$~s for the WT and $\tau_{a}=16.5\pm1.2$~s for the mutant strain, see Fig.~\ref{fig:Fig5}a. 
Regarding the time lag $\tau_{_\mathrm{delay}}$, we find that the values of the mutant strain are systematically larger than corresponding values of the WT strain, see Fig.~\ref{fig:Fig5}b. 
We hypothesize that such enhanced delay times might be caused by the absence of ChR1 and ChR2 affecting the transport of adhesion-mediating FMG-1B along the flagella. 
ChR1 and ChR2 are essential building-blocks of a signal-transduction pathway characterized by light-regulated \textit{Ca$^{2+}$}  currents that occur in the cell body (specifically at the eyespot, where ChR1 and ChR2 are localized\cite{hegemann2008algal}) and the flagellar membrane\cite{harz1991rhodopsin}. 
It is known, that \textit{Ca$^{2+}$} currents might also regulate the transport of the adhesive glycoproteins FMG-1B from the cell body to the flagella and vice versa\cite{collingridge2013compartmentalized, fort2021ca2+}. 
Thus, the deletion of ChR1 and ChR2 could potentially cause a disruption of this signalling pathway and delay the transport of FMG-1B to the flagella. 
%A second hypothesis is that the absence of the photoreceptors ChR1 and ChR2, which are responsible for controlling phototaxis\cite{berthold2008channelrhodopsin}, might prevent the cells from optimizing their perception of light (intensity), which could also result in a delayed transition to the flagellar adhesiveness. 

We also performed experiments with intensities within the threshold range (not shown) and found that a small number of cells can still switch to the adhesive state and adhere to the bottom surface. 
However, the corresponding curve do not resemble the typical sigmoidal shape and, hence, cannot be fitted using the extended Langmuir model.
So far, the focus was on the kinetics and the timescales involved in the adsorption and desorption of \textit{C.~reinhardtii} at surfaces. 
In the following section, we will now consider the density parameter $\sigma_{_{0}}$ after having established a plateau of the cell density at sufficiently long times after switching from red to blue light.

\subsection{Effect of phototactic response}

\begin{figure}[ht]
    \centering
    \includegraphics{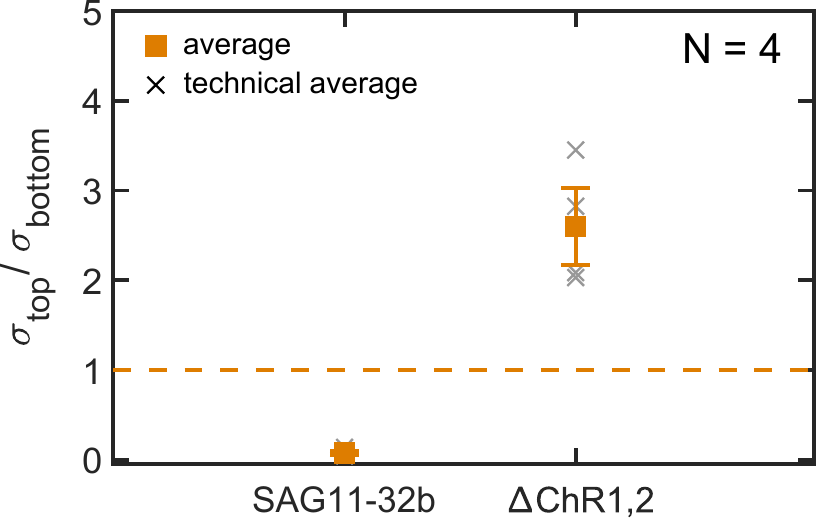}
    \caption{Comparison of the ratio of the plateau cells densities at the top and bottom surfaces achieved in four independent experiments for each strain. Knockout cells lacking the ChR1 and ChR2 photoreceptors responsible for phototaxis adsorb preferably on the top surface of the compartment. The wild-type strain SAG11-32b, however, preferably absorbs at the bottom surface. Note that the sample chamber is illuminated from the top, see Fig.~\ref{fig:setup}.}
    \label{fig:density_ratio}
\end{figure}

After changing the light conditions from red to blue light, WT cells typically exhibit photophobic and phototactic responses.
We find that at a light intensity of $I=1\cdot10^{19}$~photons$\cdot$m$^{-2}\cdot$s$^{-1}$ the initial response of the WT starts as a transient positive phototaxis, by which the cells in the suspension swim predominantly upwards in the experimental compartment.
After about 3~s, the cells then reverse to negative phototaxis and swim towards the bottom surface, where they start to adsorb after around 10~s of exposure to blue light, see Fig.~\ref{fig:Fig3}a.
As mentioned, the photoreceptors responsible for phototaxis are ChR1 and ChR2, which predominately absorb light in the blue spectrum, with maximal sensitivities at 500 and 470\,nm, respectively, and minimal absorption in the red\cite{schneider2015biophysics}. 
As a result, the strength of phototaxis in red light is negligible and we studied phototaxis only in blue/green light. 
For this set of experiments, we fixed the light intensity to $I=1\cdot10^{19}$~photons$\cdot$m$^{-2}\cdot$s$^{-1}$, and analyzed the ratio of the cell surface density measured at the top and the bottom surfaces of the compartment after the adsorption plateau is reached. 
We find that the vast majority of the WT cells adhere to the bottom boundary, see Fig.~\ref{fig:density_ratio}.
 
Typically, the cell suspensions exhibit a density of about 5$\cdot10{^6}$~cells/mL. 
If every WT cell transitions from the swimming to the surface-associated state at the bottom surface, we expect to achieve a maximum surface density between 1500-2000~cells/mm${^2}$, which is confirmed by our measurements as shown in Fig.~\ref{fig:Fig3}. 
The remarkable asymmetric distribution of adsorbed WT cells in the compartment, along with the evidence of the phototactic response, further motivated the use of channelrhodopsin-deletion mutants, namely $\Delta$ChR1,2\cite{greiner2017targeting}, for which phototactic responses in \textit{C.~reinhardtii} are effectively inhibited\cite{sineshchekov2002}.
%Regarding the time parameters, we compare the timescales shown by the mutant strain with the ones shown by the WT, see Fig.~\ref{fig:Fig4}, and note that the adsorption and desorption exhibited by the $\Delta$ChR1,2 strain is similar, although slightly slower, to the ones exhibited by WT cells, as shown in Fig.~\ref{fig:Fig4}b.
Unlike the WT, we do not find a measurable change in the direction of motion of the mutant cells after switching from red to blue light illumination, indicating that there is no photophobic or phototactic responses in the mutant strain. 
This is consistent with the fact that ChR1 and ChR2 are the main photoreceptors regulating phototaxis\cite{berthold2008channelrhodopsin}. 
Contrary to the WT individuals, the mutant cells adhere mostly at the top surface of the compartment, see Fig.~\ref{fig:density_ratio}). 
We attribute this to the fact that photoresponses are absent and cannot counteract the natural negative gravitaxis of the cells, which originates from their bottom heaviness\cite{bean1977geotactic,kam1999gravitaxis}. 
This increases the population of cells swimming near the upper part of the compartment which, after the change to blue-light conditions, increases the probability of cells adhering onto the top rather than at the bottom surface of the compartment. 

In conclusion, the invariance of the timescales exhibited by non-phototactic and phototactic individuals show that our methodology can be used as an assay for probing the ability of motile photosynthetic microorganisms to colonize surfaces. 
In addition, the parameter accounting for the ratio of plateau densities $\sigma_{_\mathrm{top}}/\sigma_{_\mathrm{bottom}}$ is governed by the strain's phototactic response and could thus be used to quantify the interplay of phototaxis and gravitaxis.\\

\section{Conclusion}

In this work we established a versatile methodology to study the kinetics of light-switchable adsorption of \textit{C.~reinhardtii} based on a time-delayed Langmuir model for microbial adsorption and desorption at solid surfaces. 
We show that both adsorption and desorption exhibit a lag response relative to the time at which the blue- or red-light conditions are set. 
After exposure to blue light, the cells adhere to the surface with a characteristic delay time in the order of 10s of seconds, whereas for the desorption the timescale is around one order of magnitude larger.
%50\% of the cells become sticky within 16 - 20~s, whereas for the desorption the timescale is around one order of magnitude longer.
This delay time in the adsorption decreases with increasing light intensity, however the rate at which the cells adsorb is independent of the light intensity. 
The adsorption and desorption kinetics exhibited in channelrhodopsin-deficient cells is comparable to the wild-type individuals, hence phototaxis does not significantly affect the adsorption rate.
%The delay time shows then that light-switchable adhesion is a very slow process compared to other light-induced responses (e.g. photophobic and phototactic), which typically occur in a timescale of tens of milliseconds \cite{harz1991rhodopsin}.
It has not escaped our attention that the photoreceptors ChR1 and ChR2 are not the ones responsible for light-switchable adhesion in \textit{C.~reinhardtii}, hence future research has to focus on other photoreceptors in order to elucidate the mechanism of this particular trait. 
Finally, we state that the invariance in the timescales of adsorption and desorption between the blind and the WT strains allows our methodology to be used as an assay for surface colonization of photoactive microorganisms.

\section*{Conflicts of interest}
There are no conflicts to declare.\\
%In accordance with our policy on \href{http://www.rsc.org/journals-books-databases/journal-authors-reviewers/author-responsibilities/#code-of-conduct}{Conflicts of interest} please ensure that a conflicts of interest statement is included in your manuscript here.  Please note that this statement is required for all submitted manuscripts.  If no conflicts exist, please state that ``There are no conflicts to declare''.

\section*{Acknowledgements}
The authors thank the Göttingen Algae Culture Collection (SAG) for providing the \textit{C.~reinhardtii} strain SAG11-32b and M.~Lorenz for discussions and technical assistance. 
We thank H.~Evers and F.~Böhning for their excellent technical assistance in generating the $\Delta$ChR1,2 mutant and C.T.~Kreis, M.~Müller, J.~Enderlein and S.~Herminghaus for stimulating discussions. 
R.~Catalan acknowledges the German Academic Exchange Service (DAAD) for generous financial support.\\
%The Acknowledgements come at the end of an article after Conflicts of interest and before the Notes and references.

%


\begin{thebibliography}{77}%
\makeatletter
\providecommand \@ifxundefined [1]{%
 \@ifx{#1\undefined}
}%
\providecommand \@ifnum [1]{%
 \ifnum #1\expandafter \@firstoftwo
 \else \expandafter \@secondoftwo
 \fi
}%
\providecommand \@ifx [1]{%
 \ifx #1\expandafter \@firstoftwo
 \else \expandafter \@secondoftwo
 \fi
}%
\providecommand \natexlab [1]{#1}%
\providecommand \enquote  [1]{``#1''}%
\providecommand \bibnamefont  [1]{#1}%
\providecommand \bibfnamefont [1]{#1}%
\providecommand \citenamefont [1]{#1}%
\providecommand \href@noop [0]{\@secondoftwo}%
\providecommand \href [0]{\begingroup \@sanitize@url \@href}%
\providecommand \@href[1]{\@@startlink{#1}\@@href}%
\providecommand \@@href[1]{\endgroup#1\@@endlink}%
\providecommand \@sanitize@url [0]{\catcode `\\12\catcode `\$12\catcode
  `\&12\catcode `\#12\catcode `\^12\catcode `\_12\catcode `\%12\relax}%
\providecommand \@@startlink[1]{}%
\providecommand \@@endlink[0]{}%
\providecommand \url  [0]{\begingroup\@sanitize@url \@url }%
\providecommand \@url [1]{\endgroup\@href {#1}{\urlprefix }}%
\providecommand \urlprefix  [0]{URL }%
\providecommand \Eprint [0]{\href }%
\providecommand \doibase [0]{https://doi.org/}%
\providecommand \selectlanguage [0]{\@gobble}%
\providecommand \bibinfo  [0]{\@secondoftwo}%
\providecommand \bibfield  [0]{\@secondoftwo}%
\providecommand \translation [1]{[#1]}%
\providecommand \BibitemOpen [0]{}%
\providecommand \bibitemStop [0]{}%
\providecommand \bibitemNoStop [0]{.\EOS\space}%
\providecommand \EOS [0]{\spacefactor3000\relax}%
\providecommand \BibitemShut  [1]{\csname bibitem#1\endcsname}%
\let\auto@bib@innerbib\@empty
%</preamble>
\bibitem [{\citenamefont {Noffke}\ \emph {et~al.}(2013)\citenamefont {Noffke},
  \citenamefont {Christian}, \citenamefont {Wacey},\ and\ \citenamefont
  {Hazen}}]{noffke2013microbially}%
  \BibitemOpen
  \bibfield  {author} {\bibinfo {author} {\bibfnamefont {N.}~\bibnamefont
  {Noffke}}, \bibinfo {author} {\bibfnamefont {D.}~\bibnamefont {Christian}},
  \bibinfo {author} {\bibfnamefont {D.}~\bibnamefont {Wacey}},\ and\ \bibinfo
  {author} {\bibfnamefont {R.~M.}\ \bibnamefont {Hazen}},\ }\bibfield  {title}
  {\bibinfo {title} {{M}icrobially induced sedimentary structures recording an
  ancient ecosystem in the ca. 3.48 billion-year-old {D}resser {F}ormation,
  {P}ilbara, {W}estern {A}ustralia},\ }\href
  {https://doi.org/10.1089/ast.2013.1030} {\bibfield  {journal} {\bibinfo
  {journal} {Astrobiology}\ }\textbf {\bibinfo {volume} {13}},\ \bibinfo
  {pages} {1103} (\bibinfo {year} {2013})}\BibitemShut {NoStop}%
\bibitem [{\citenamefont {Rasmussen}(2000)}]{rasmussen2000filamentous}%
  \BibitemOpen
  \bibfield  {author} {\bibinfo {author} {\bibfnamefont {B.}~\bibnamefont
  {Rasmussen}},\ }\bibfield  {title} {\bibinfo {title} {Filamentous
  microfossils in a 3,235-million-year-old volcanogenic massive sulphide
  deposit},\ }\href@noop {} {\bibfield  {journal} {\bibinfo  {journal}
  {Nature}\ }\textbf {\bibinfo {volume} {405}},\ \bibinfo {pages} {676}
  (\bibinfo {year} {2000})}\BibitemShut {NoStop}%
\bibitem [{\citenamefont {Li}\ \emph {et~al.}(2004)\citenamefont {Li},
  \citenamefont {Helmerhorst}, \citenamefont {Leone}, \citenamefont {Troxler},
  \citenamefont {Yaskell}, \citenamefont {Haffajee}, \citenamefont
  {Socransky},\ and\ \citenamefont {Oppenheim}}]{li2004identification}%
  \BibitemOpen
  \bibfield  {author} {\bibinfo {author} {\bibfnamefont {J.}~\bibnamefont
  {Li}}, \bibinfo {author} {\bibfnamefont {E.~J.}\ \bibnamefont {Helmerhorst}},
  \bibinfo {author} {\bibfnamefont {C.~W.}\ \bibnamefont {Leone}}, \bibinfo
  {author} {\bibfnamefont {R.~F.}\ \bibnamefont {Troxler}}, \bibinfo {author}
  {\bibfnamefont {T.}~\bibnamefont {Yaskell}}, \bibinfo {author} {\bibfnamefont
  {A.~D.}\ \bibnamefont {Haffajee}}, \bibinfo {author} {\bibfnamefont {S.~S.}\
  \bibnamefont {Socransky}},\ and\ \bibinfo {author} {\bibfnamefont
  {F.}~\bibnamefont {Oppenheim}},\ }\bibfield  {title} {\bibinfo {title}
  {Identification of early microbial colonizers in human dental biofilm},\
  }\href@noop {} {\bibfield  {journal} {\bibinfo  {journal} {J. Appl.
  Microbiol.}\ }\textbf {\bibinfo {volume} {97}},\ \bibinfo {pages} {1311}
  (\bibinfo {year} {2004})}\BibitemShut {NoStop}%
\bibitem [{\citenamefont {Motta}\ \emph {et~al.}(2021)\citenamefont {Motta},
  \citenamefont {Wallace}, \citenamefont {Buret}, \citenamefont {Deraison},\
  and\ \citenamefont {Vergnolle}}]{motta2021gastrointestinal}%
  \BibitemOpen
  \bibfield  {author} {\bibinfo {author} {\bibfnamefont {J.-P.}\ \bibnamefont
  {Motta}}, \bibinfo {author} {\bibfnamefont {J.~L.}\ \bibnamefont {Wallace}},
  \bibinfo {author} {\bibfnamefont {A.~G.}\ \bibnamefont {Buret}}, \bibinfo
  {author} {\bibfnamefont {C.}~\bibnamefont {Deraison}},\ and\ \bibinfo
  {author} {\bibfnamefont {N.}~\bibnamefont {Vergnolle}},\ }\bibfield  {title}
  {\bibinfo {title} {Gastrointestinal biofilms in health and disease},\
  }\href@noop {} {\bibfield  {journal} {\bibinfo  {journal} {Nature Reviews
  Gastroenterology \& Hepatology}\ }\textbf {\bibinfo {volume} {18}},\ \bibinfo
  {pages} {314} (\bibinfo {year} {2021})}\BibitemShut {NoStop}%
\bibitem [{\citenamefont {Stickler}(2008)}]{stickler2008bacterial}%
  \BibitemOpen
  \bibfield  {author} {\bibinfo {author} {\bibfnamefont {D.~J.}\ \bibnamefont
  {Stickler}},\ }\bibfield  {title} {\bibinfo {title} {Bacterial biofilms in
  patients with indwelling urinary catheters},\ }\href@noop {} {\bibfield
  {journal} {\bibinfo  {journal} {Nature clinical practice urology}\ }\textbf
  {\bibinfo {volume} {5}},\ \bibinfo {pages} {598} (\bibinfo {year}
  {2008})}\BibitemShut {NoStop}%
\bibitem [{\citenamefont {Flemming}\ \emph {et~al.}(2016)\citenamefont
  {Flemming}, \citenamefont {Wingender}, \citenamefont {Szewzyk}, \citenamefont
  {Steinberg}, \citenamefont {Rice},\ and\ \citenamefont
  {Kjelleberg}}]{flemming2016biofilms}%
  \BibitemOpen
  \bibfield  {author} {\bibinfo {author} {\bibfnamefont {H.-C.}\ \bibnamefont
  {Flemming}}, \bibinfo {author} {\bibfnamefont {J.}~\bibnamefont {Wingender}},
  \bibinfo {author} {\bibfnamefont {U.}~\bibnamefont {Szewzyk}}, \bibinfo
  {author} {\bibfnamefont {P.}~\bibnamefont {Steinberg}}, \bibinfo {author}
  {\bibfnamefont {S.~A.}\ \bibnamefont {Rice}},\ and\ \bibinfo {author}
  {\bibfnamefont {S.}~\bibnamefont {Kjelleberg}},\ }\bibfield  {title}
  {\bibinfo {title} {Biofilms: an emergent form of bacterial life},\
  }\href@noop {} {\bibfield  {journal} {\bibinfo  {journal} {Nat. Rev.
  Microbiol.}\ }\textbf {\bibinfo {volume} {14}},\ \bibinfo {pages} {563}
  (\bibinfo {year} {2016})}\BibitemShut {NoStop}%
\bibitem [{\citenamefont {O'Toole}\ \emph {et~al.}(2000)\citenamefont
  {O'Toole}, \citenamefont {Kaplan},\ and\ \citenamefont
  {Kolter}}]{o2000biofilm}%
  \BibitemOpen
  \bibfield  {author} {\bibinfo {author} {\bibfnamefont {G.}~\bibnamefont
  {O'Toole}}, \bibinfo {author} {\bibfnamefont {H.~B.}\ \bibnamefont
  {Kaplan}},\ and\ \bibinfo {author} {\bibfnamefont {R.}~\bibnamefont
  {Kolter}},\ }\bibfield  {title} {\bibinfo {title} {Biofilm formation as
  microbial development},\ }\href@noop {} {\bibfield  {journal} {\bibinfo
  {journal} {Annu. Rev. Microbiol.}\ }\textbf {\bibinfo {volume} {54}},\
  \bibinfo {pages} {49} (\bibinfo {year} {2000})}\BibitemShut {NoStop}%
\bibitem [{\citenamefont {Stewart}\ and\ \citenamefont
  {Costerton}(2001)}]{stewart2001antibiotic}%
  \BibitemOpen
  \bibfield  {author} {\bibinfo {author} {\bibfnamefont {P.~S.}\ \bibnamefont
  {Stewart}}\ and\ \bibinfo {author} {\bibfnamefont {J.~W.}\ \bibnamefont
  {Costerton}},\ }\bibfield  {title} {\bibinfo {title} {Antibiotic resistance
  of bacteria in biofilms},\ }\href@noop {} {\bibfield  {journal} {\bibinfo
  {journal} {Lancet}\ }\textbf {\bibinfo {volume} {358}},\ \bibinfo {pages}
  {135} (\bibinfo {year} {2001})}\BibitemShut {NoStop}%
\bibitem [{\citenamefont {Mah}\ and\ \citenamefont
  {O'Toole}(2001)}]{mah2001mechanisms}%
  \BibitemOpen
  \bibfield  {author} {\bibinfo {author} {\bibfnamefont {T.-F.~C.}\
  \bibnamefont {Mah}}\ and\ \bibinfo {author} {\bibfnamefont {G.~A.}\
  \bibnamefont {O'Toole}},\ }\bibfield  {title} {\bibinfo {title} {Mechanisms
  of biofilm resistance to antimicrobial agents},\ }\href@noop {} {\bibfield
  {journal} {\bibinfo  {journal} {Trends Microbiol.}\ }\textbf {\bibinfo
  {volume} {9}},\ \bibinfo {pages} {34} (\bibinfo {year} {2001})}\BibitemShut
  {NoStop}%
\bibitem [{\citenamefont {Bridier}\ \emph {et~al.}(2011)\citenamefont
  {Bridier}, \citenamefont {Briandet}, \citenamefont {Thomas},\ and\
  \citenamefont {Dubois-Brissonnet}}]{bridier2011resistance}%
  \BibitemOpen
  \bibfield  {author} {\bibinfo {author} {\bibfnamefont {A.}~\bibnamefont
  {Bridier}}, \bibinfo {author} {\bibfnamefont {R.}~\bibnamefont {Briandet}},
  \bibinfo {author} {\bibfnamefont {V.}~\bibnamefont {Thomas}},\ and\ \bibinfo
  {author} {\bibfnamefont {F.}~\bibnamefont {Dubois-Brissonnet}},\ }\bibfield
  {title} {\bibinfo {title} {Resistance of bacterial biofilms to disinfectants:
  a review},\ }\href@noop {} {\bibfield  {journal} {\bibinfo  {journal}
  {Biofouling}\ }\textbf {\bibinfo {volume} {27}},\ \bibinfo {pages} {1017}
  (\bibinfo {year} {2011})}\BibitemShut {NoStop}%
\bibitem [{\citenamefont {Hall-Stoodley}\ \emph {et~al.}(2004)\citenamefont
  {Hall-Stoodley}, \citenamefont {Costerton},\ and\ \citenamefont
  {Stoodley}}]{hall2004bacterial}%
  \BibitemOpen
  \bibfield  {author} {\bibinfo {author} {\bibfnamefont {L.}~\bibnamefont
  {Hall-Stoodley}}, \bibinfo {author} {\bibfnamefont {J.~W.}\ \bibnamefont
  {Costerton}},\ and\ \bibinfo {author} {\bibfnamefont {P.}~\bibnamefont
  {Stoodley}},\ }\bibfield  {title} {\bibinfo {title} {Bacterial biofilms: from
  the natural environment to infectious diseases},\ }\href@noop {} {\bibfield
  {journal} {\bibinfo  {journal} {Nat. Rev. Microbiol.}\ }\textbf {\bibinfo
  {volume} {2}},\ \bibinfo {pages} {95} (\bibinfo {year} {2004})}\BibitemShut
  {NoStop}%
\bibitem [{\citenamefont {Costerton}\ \emph {et~al.}(1999)\citenamefont
  {Costerton}, \citenamefont {Stewart},\ and\ \citenamefont
  {Greenberg}}]{costerton1999}%
  \BibitemOpen
  \bibfield  {author} {\bibinfo {author} {\bibfnamefont {J.~W.}\ \bibnamefont
  {Costerton}}, \bibinfo {author} {\bibfnamefont {P.~S.}\ \bibnamefont
  {Stewart}},\ and\ \bibinfo {author} {\bibfnamefont {E.~P.}\ \bibnamefont
  {Greenberg}},\ }\bibfield  {title} {\bibinfo {title} {Bacterial biofilms: A
  common cause of persistent infections},\ }\href
  {https://doi.org/10.1126/science.284.5418.1318} {\bibfield  {journal}
  {\bibinfo  {journal} {Science}\ }\textbf {\bibinfo {volume} {284}},\ \bibinfo
  {pages} {1318} (\bibinfo {year} {1999})},\ \Eprint
  {https://arxiv.org/abs/https://www.science.org/doi/pdf/10.1126/science.284.5418.1318}
  {https://www.science.org/doi/pdf/10.1126/science.284.5418.1318} \BibitemShut
  {NoStop}%
\bibitem [{\citenamefont {Nicolella}\ \emph {et~al.}(2000)\citenamefont
  {Nicolella}, \citenamefont {van Loosdrecht},\ and\ \citenamefont
  {Heijnen}}]{nicolella2000}%
  \BibitemOpen
  \bibfield  {author} {\bibinfo {author} {\bibfnamefont {C.}~\bibnamefont
  {Nicolella}}, \bibinfo {author} {\bibfnamefont {M.~C.}\ \bibnamefont {van
  Loosdrecht}},\ and\ \bibinfo {author} {\bibfnamefont {S.~J.}\ \bibnamefont
  {Heijnen}},\ }\bibfield  {title} {\bibinfo {title} {Particle-based biofilm
  reactor technology},\ }\href@noop {} {\bibfield  {journal} {\bibinfo
  {journal} {Trends Biotechnol.}\ }\textbf {\bibinfo {volume} {18}},\ \bibinfo
  {pages} {312} (\bibinfo {year} {2000})}\BibitemShut {NoStop}%
\bibitem [{\citenamefont {Van~Loosdrecht}\ and\ \citenamefont
  {Heijnen}(1993)}]{van1993}%
  \BibitemOpen
  \bibfield  {author} {\bibinfo {author} {\bibfnamefont {M.~C.}\ \bibnamefont
  {Van~Loosdrecht}}\ and\ \bibinfo {author} {\bibfnamefont {S.~J.}\
  \bibnamefont {Heijnen}},\ }\bibfield  {title} {\bibinfo {title} {Biofilm
  bioreactors for waste-water treatment},\ }\href@noop {} {\bibfield  {journal}
  {\bibinfo  {journal} {Trends Biotechnol.}\ }\textbf {\bibinfo {volume}
  {11}},\ \bibinfo {pages} {117} (\bibinfo {year} {1993})}\BibitemShut
  {NoStop}%
\bibitem [{\citenamefont {Kumar}\ and\ \citenamefont
  {Anand}(1998)}]{kumar1998}%
  \BibitemOpen
  \bibfield  {author} {\bibinfo {author} {\bibfnamefont {C.~G.}\ \bibnamefont
  {Kumar}}\ and\ \bibinfo {author} {\bibfnamefont {S.}~\bibnamefont {Anand}},\
  }\bibfield  {title} {\bibinfo {title} {Significance of microbial biofilms in
  food industry: a review},\ }\href@noop {} {\bibfield  {journal} {\bibinfo
  {journal} {Int. J. Food Microbiol.}\ }\textbf {\bibinfo {volume} {42}},\
  \bibinfo {pages} {9} (\bibinfo {year} {1998})}\BibitemShut {NoStop}%
\bibitem [{\citenamefont {Francolini}\ and\ \citenamefont
  {Donelli}(2010)}]{francolini2010prevention}%
  \BibitemOpen
  \bibfield  {author} {\bibinfo {author} {\bibfnamefont {I.}~\bibnamefont
  {Francolini}}\ and\ \bibinfo {author} {\bibfnamefont {G.}~\bibnamefont
  {Donelli}},\ }\bibfield  {title} {\bibinfo {title} {Prevention and control of
  biofilm-based medical-device-related infections},\ }\href@noop {} {\bibfield
  {journal} {\bibinfo  {journal} {FEMS Immunology \& Medical Microbiology}\
  }\textbf {\bibinfo {volume} {59}},\ \bibinfo {pages} {227} (\bibinfo {year}
  {2010})}\BibitemShut {NoStop}%
\bibitem [{\citenamefont {Thompson}\ \emph {et~al.}(2002)\citenamefont
  {Thompson}, \citenamefont {Abreu},\ and\ \citenamefont
  {Wasielesky}}]{thompson2002importance}%
  \BibitemOpen
  \bibfield  {author} {\bibinfo {author} {\bibfnamefont {F.~L.}\ \bibnamefont
  {Thompson}}, \bibinfo {author} {\bibfnamefont {P.~C.}\ \bibnamefont
  {Abreu}},\ and\ \bibinfo {author} {\bibfnamefont {W.}~\bibnamefont
  {Wasielesky}},\ }\bibfield  {title} {\bibinfo {title} {Importance of biofilm
  for water quality and nourishment in intensive shrimp culture},\ }\href@noop
  {} {\bibfield  {journal} {\bibinfo  {journal} {Aquaculture}\ }\textbf
  {\bibinfo {volume} {203}},\ \bibinfo {pages} {263} (\bibinfo {year}
  {2002})}\BibitemShut {NoStop}%
\bibitem [{\citenamefont {Armbruster}\ and\ \citenamefont
  {Parsek}(2018)}]{armbruster2018}%
  \BibitemOpen
  \bibfield  {author} {\bibinfo {author} {\bibfnamefont {C.~R.}\ \bibnamefont
  {Armbruster}}\ and\ \bibinfo {author} {\bibfnamefont {M.~R.}\ \bibnamefont
  {Parsek}},\ }\bibfield  {title} {\bibinfo {title} {New insight into the early
  stages of biofilm formation},\ }\href@noop {} {\bibfield  {journal} {\bibinfo
   {journal} {Proc. Natl. Acad. Sci. U. S. A.}\ }\textbf {\bibinfo {volume}
  {115}},\ \bibinfo {pages} {4317} (\bibinfo {year} {2018})}\BibitemShut
  {NoStop}%
\bibitem [{\citenamefont {Utada}\ \emph {et~al.}(2014)\citenamefont {Utada},
  \citenamefont {Bennett}, \citenamefont {Fong}, \citenamefont {Gibiansky},
  \citenamefont {Yildiz}, \citenamefont {Golestanian},\ and\ \citenamefont
  {Wong}}]{utada2014vibrio}%
  \BibitemOpen
  \bibfield  {author} {\bibinfo {author} {\bibfnamefont {A.~S.}\ \bibnamefont
  {Utada}}, \bibinfo {author} {\bibfnamefont {R.~R.}\ \bibnamefont {Bennett}},
  \bibinfo {author} {\bibfnamefont {J.~C.}\ \bibnamefont {Fong}}, \bibinfo
  {author} {\bibfnamefont {M.~L.}\ \bibnamefont {Gibiansky}}, \bibinfo {author}
  {\bibfnamefont {F.~H.}\ \bibnamefont {Yildiz}}, \bibinfo {author}
  {\bibfnamefont {R.}~\bibnamefont {Golestanian}},\ and\ \bibinfo {author}
  {\bibfnamefont {G.~C.}\ \bibnamefont {Wong}},\ }\bibfield  {title} {\bibinfo
  {title} {Vibrio cholerae use pili and flagella synergistically to effect
  motility switching and conditional surface attachment},\ }\href@noop {}
  {\bibfield  {journal} {\bibinfo  {journal} {Nature communications}\ }\textbf
  {\bibinfo {volume} {5}},\ \bibinfo {pages} {1} (\bibinfo {year}
  {2014})}\BibitemShut {NoStop}%
\bibitem [{\citenamefont {Marshall}\ \emph {et~al.}(1971)\citenamefont
  {Marshall}, \citenamefont {Stout},\ and\ \citenamefont
  {Mitchell}}]{marshall1971mechanism}%
  \BibitemOpen
  \bibfield  {author} {\bibinfo {author} {\bibfnamefont {K.}~\bibnamefont
  {Marshall}}, \bibinfo {author} {\bibfnamefont {R.}~\bibnamefont {Stout}},\
  and\ \bibinfo {author} {\bibfnamefont {R.}~\bibnamefont {Mitchell}},\
  }\bibfield  {title} {\bibinfo {title} {Mechanism of the initial events in the
  sorption of marine bacteria to surfaces},\ }\href@noop {} {\bibfield
  {journal} {\bibinfo  {journal} {Microbiology}\ }\textbf {\bibinfo {volume}
  {68}},\ \bibinfo {pages} {337} (\bibinfo {year} {1971})}\BibitemShut
  {NoStop}%
\bibitem [{\citenamefont {Loskill}\ \emph {et~al.}(2012)\citenamefont
  {Loskill}, \citenamefont {Hähl}, \citenamefont {Thewes}, \citenamefont
  {Kreis}, \citenamefont {Bischoff}, \citenamefont {Herrmann},\ and\
  \citenamefont {Jacobs}}]{loskill2012}%
  \BibitemOpen
  \bibfield  {author} {\bibinfo {author} {\bibfnamefont {P.}~\bibnamefont
  {Loskill}}, \bibinfo {author} {\bibfnamefont {H.}~\bibnamefont {Hähl}},
  \bibinfo {author} {\bibfnamefont {N.}~\bibnamefont {Thewes}}, \bibinfo
  {author} {\bibfnamefont {C.~T.}\ \bibnamefont {Kreis}}, \bibinfo {author}
  {\bibfnamefont {M.}~\bibnamefont {Bischoff}}, \bibinfo {author}
  {\bibfnamefont {M.}~\bibnamefont {Herrmann}},\ and\ \bibinfo {author}
  {\bibfnamefont {K.}~\bibnamefont {Jacobs}},\ }\bibfield  {title} {\bibinfo
  {title} {Influence of the subsurface composition of a material on the
  adhesion of staphylococci},\ }\href {https://doi.org/10.1021/la3004323}
  {\bibfield  {journal} {\bibinfo  {journal} {Langmuir}\ }\textbf {\bibinfo
  {volume} {28}},\ \bibinfo {pages} {7242} (\bibinfo {year} {2012})},\ \bibinfo
  {note} {pMID: 22475009},\ \Eprint
  {https://arxiv.org/abs/https://doi.org/10.1021/la3004323}
  {https://doi.org/10.1021/la3004323} \BibitemShut {NoStop}%
\bibitem [{\citenamefont {Maikranz}\ \emph {et~al.}(2020)\citenamefont
  {Maikranz}, \citenamefont {Spengler}, \citenamefont {Thewes}, \citenamefont
  {Thewes}, \citenamefont {Nolle}, \citenamefont {Jung}, \citenamefont
  {Bischoff}, \citenamefont {Santen},\ and\ \citenamefont
  {Jacobs}}]{maikranz2020different}%
  \BibitemOpen
  \bibfield  {author} {\bibinfo {author} {\bibfnamefont {E.}~\bibnamefont
  {Maikranz}}, \bibinfo {author} {\bibfnamefont {C.}~\bibnamefont {Spengler}},
  \bibinfo {author} {\bibfnamefont {N.}~\bibnamefont {Thewes}}, \bibinfo
  {author} {\bibfnamefont {A.}~\bibnamefont {Thewes}}, \bibinfo {author}
  {\bibfnamefont {F.}~\bibnamefont {Nolle}}, \bibinfo {author} {\bibfnamefont
  {P.}~\bibnamefont {Jung}}, \bibinfo {author} {\bibfnamefont {M.}~\bibnamefont
  {Bischoff}}, \bibinfo {author} {\bibfnamefont {L.}~\bibnamefont {Santen}},\
  and\ \bibinfo {author} {\bibfnamefont {K.}~\bibnamefont {Jacobs}},\
  }\bibfield  {title} {\bibinfo {title} {Different binding mechanisms of
  staphylococcus aureus to hydrophobic and hydrophilic surfaces},\ }\href@noop
  {} {\bibfield  {journal} {\bibinfo  {journal} {Nanoscale}\ }\textbf {\bibinfo
  {volume} {12}},\ \bibinfo {pages} {19267} (\bibinfo {year}
  {2020})}\BibitemShut {NoStop}%
\bibitem [{\citenamefont {Dufr{\^e}ne}\ and\ \citenamefont
  {Persat}(2020)}]{dufrene2020mechanomicrobiology}%
  \BibitemOpen
  \bibfield  {author} {\bibinfo {author} {\bibfnamefont {Y.~F.}\ \bibnamefont
  {Dufr{\^e}ne}}\ and\ \bibinfo {author} {\bibfnamefont {A.}~\bibnamefont
  {Persat}},\ }\bibfield  {title} {\bibinfo {title} {Mechanomicrobiology: how
  bacteria sense and respond to forces},\ }\href@noop {} {\bibfield  {journal}
  {\bibinfo  {journal} {Nature Reviews Microbiology}\ }\textbf {\bibinfo
  {volume} {18}},\ \bibinfo {pages} {227} (\bibinfo {year} {2020})}\BibitemShut
  {NoStop}%
\bibitem [{\citenamefont {Marshall}(2006)}]{marshall2006}%
  \BibitemOpen
  \bibfield  {author} {\bibinfo {author} {\bibfnamefont {K.~C.}\ \bibnamefont
  {Marshall}},\ }\bibfield  {title} {\bibinfo {title} {Planktonic versus
  sessile life of prokaryotes},\ }\href@noop {} {\bibfield  {journal} {\bibinfo
   {journal} {The prokaryotes}\ }\textbf {\bibinfo {volume} {2}},\ \bibinfo
  {pages} {3} (\bibinfo {year} {2006})}\BibitemShut {NoStop}%
\bibitem [{\citenamefont {D{\'\i}az}\ \emph {et~al.}(2007)\citenamefont
  {D{\'\i}az}, \citenamefont {Schilardi}, \citenamefont {Salvarezza},\ and\
  \citenamefont {Fern{\'a}ndez Lorenzo~de Mele}}]{diaz2007}%
  \BibitemOpen
  \bibfield  {author} {\bibinfo {author} {\bibfnamefont {C.}~\bibnamefont
  {D{\'\i}az}}, \bibinfo {author} {\bibfnamefont {P.}~\bibnamefont
  {Schilardi}}, \bibinfo {author} {\bibfnamefont {R.}~\bibnamefont
  {Salvarezza}},\ and\ \bibinfo {author} {\bibfnamefont {M.}~\bibnamefont
  {Fern{\'a}ndez Lorenzo~de Mele}},\ }\bibfield  {title} {\bibinfo {title}
  {Nano/microscale order affects the early stages of biofilm formation on metal
  surfaces},\ }\href@noop {} {\bibfield  {journal} {\bibinfo  {journal}
  {Langmuir}\ }\textbf {\bibinfo {volume} {23}},\ \bibinfo {pages} {11206}
  (\bibinfo {year} {2007})}\BibitemShut {NoStop}%
\bibitem [{\citenamefont {Sun}\ \emph {et~al.}(2020)\citenamefont {Sun},
  \citenamefont {Jarisz}, \citenamefont {Wunsch},\ and\ \citenamefont
  {Hore}}]{sun2020}%
  \BibitemOpen
  \bibfield  {author} {\bibinfo {author} {\bibfnamefont {V.}~\bibnamefont
  {Sun}}, \bibinfo {author} {\bibfnamefont {T.~A.}\ \bibnamefont {Jarisz}},
  \bibinfo {author} {\bibfnamefont {L.}~\bibnamefont {Wunsch}},\ and\ \bibinfo
  {author} {\bibfnamefont {D.~K.}\ \bibnamefont {Hore}},\ }\bibfield  {title}
  {\bibinfo {title} {Monitoring early stages of bacterial adhesion at silica
  surfaces through image analysis},\ }\href@noop {} {\bibfield  {journal}
  {\bibinfo  {journal} {Langmuir}\ }\textbf {\bibinfo {volume} {36}},\ \bibinfo
  {pages} {2120} (\bibinfo {year} {2020})}\BibitemShut {NoStop}%
\bibitem [{\citenamefont {Lee}\ \emph {et~al.}(2018)\citenamefont {Lee},
  \citenamefont {de~Anda}, \citenamefont {Baker}, \citenamefont {Bennett},
  \citenamefont {Luo}, \citenamefont {Lee}, \citenamefont {Keefe},
  \citenamefont {Helali}, \citenamefont {Ma}, \citenamefont {Zhao} \emph
  {et~al.}}]{lee2018}%
  \BibitemOpen
  \bibfield  {author} {\bibinfo {author} {\bibfnamefont {C.~K.}\ \bibnamefont
  {Lee}}, \bibinfo {author} {\bibfnamefont {J.}~\bibnamefont {de~Anda}},
  \bibinfo {author} {\bibfnamefont {A.~E.}\ \bibnamefont {Baker}}, \bibinfo
  {author} {\bibfnamefont {R.~R.}\ \bibnamefont {Bennett}}, \bibinfo {author}
  {\bibfnamefont {Y.}~\bibnamefont {Luo}}, \bibinfo {author} {\bibfnamefont
  {E.~Y.}\ \bibnamefont {Lee}}, \bibinfo {author} {\bibfnamefont {J.~A.}\
  \bibnamefont {Keefe}}, \bibinfo {author} {\bibfnamefont {J.~S.}\ \bibnamefont
  {Helali}}, \bibinfo {author} {\bibfnamefont {J.}~\bibnamefont {Ma}}, \bibinfo
  {author} {\bibfnamefont {K.}~\bibnamefont {Zhao}}, \emph {et~al.},\
  }\bibfield  {title} {\bibinfo {title} {Multigenerational memory and adaptive
  adhesion in early bacterial biofilm communities},\ }\href@noop {} {\bibfield
  {journal} {\bibinfo  {journal} {Proc. Natl. Acad. Sci. U. S. A.}\ }\textbf
  {\bibinfo {volume} {115}},\ \bibinfo {pages} {4471} (\bibinfo {year}
  {2018})}\BibitemShut {NoStop}%
\bibitem [{\citenamefont {Mazza}(2016)}]{mazza2016}%
  \BibitemOpen
  \bibfield  {author} {\bibinfo {author} {\bibfnamefont {M.~G.}\ \bibnamefont
  {Mazza}},\ }\bibfield  {title} {\bibinfo {title} {The physics of
  biofilms—an introduction},\ }\href@noop {} {\bibfield  {journal} {\bibinfo
  {journal} {J. Phys. D}\ }\textbf {\bibinfo {volume} {49}},\ \bibinfo {pages}
  {203001} (\bibinfo {year} {2016})}\BibitemShut {NoStop}%
\bibitem [{\citenamefont {Beardall}\ and\ \citenamefont
  {Giordano}(2002)}]{beardall2002ecological}%
  \BibitemOpen
  \bibfield  {author} {\bibinfo {author} {\bibfnamefont {J.}~\bibnamefont
  {Beardall}}\ and\ \bibinfo {author} {\bibfnamefont {M.}~\bibnamefont
  {Giordano}},\ }\bibfield  {title} {\bibinfo {title} {Ecological implications
  of microalgal and cyanobacterial co2 concentrating mechanisms, and their
  regulation},\ }\href@noop {} {\bibfield  {journal} {\bibinfo  {journal}
  {Functional Plant Biology}\ }\textbf {\bibinfo {volume} {29}},\ \bibinfo
  {pages} {335} (\bibinfo {year} {2002})}\BibitemShut {NoStop}%
\bibitem [{\citenamefont {Lamenti}\ \emph {et~al.}(2000)\citenamefont
  {Lamenti}, \citenamefont {Tiano},\ and\ \citenamefont
  {Tomaselli}}]{lamenti2000biodeterioration}%
  \BibitemOpen
  \bibfield  {author} {\bibinfo {author} {\bibfnamefont {G.}~\bibnamefont
  {Lamenti}}, \bibinfo {author} {\bibfnamefont {P.}~\bibnamefont {Tiano}},\
  and\ \bibinfo {author} {\bibfnamefont {L.}~\bibnamefont {Tomaselli}},\
  }\bibfield  {title} {\bibinfo {title} {Biodeterioration of ornamental marble
  statues in the boboli gardens (florence, italy)},\ }\href@noop {} {\bibfield
  {journal} {\bibinfo  {journal} {J. Appl. Phycol.}\ }\textbf {\bibinfo
  {volume} {12}},\ \bibinfo {pages} {427} (\bibinfo {year} {2000})}\BibitemShut
  {NoStop}%
\bibitem [{\citenamefont {Schultz}\ \emph {et~al.}(2011)\citenamefont
  {Schultz}, \citenamefont {Bendick}, \citenamefont {Holm},\ and\ \citenamefont
  {Hertel}}]{schultz2011economic}%
  \BibitemOpen
  \bibfield  {author} {\bibinfo {author} {\bibfnamefont {M.}~\bibnamefont
  {Schultz}}, \bibinfo {author} {\bibfnamefont {J.}~\bibnamefont {Bendick}},
  \bibinfo {author} {\bibfnamefont {E.}~\bibnamefont {Holm}},\ and\ \bibinfo
  {author} {\bibfnamefont {W.}~\bibnamefont {Hertel}},\ }\bibfield  {title}
  {\bibinfo {title} {Economic impact of biofouling on a naval surface ship},\
  }\href@noop {} {\bibfield  {journal} {\bibinfo  {journal} {Biofouling}\
  }\textbf {\bibinfo {volume} {27}},\ \bibinfo {pages} {87} (\bibinfo {year}
  {2011})}\BibitemShut {NoStop}%
\bibitem [{\citenamefont {Boelee}\ \emph {et~al.}(2011)\citenamefont {Boelee},
  \citenamefont {Temmink}, \citenamefont {Janssen}, \citenamefont {Buisman},\
  and\ \citenamefont {Wijffels}}]{boelee2011}%
  \BibitemOpen
  \bibfield  {author} {\bibinfo {author} {\bibfnamefont {N.}~\bibnamefont
  {Boelee}}, \bibinfo {author} {\bibfnamefont {H.}~\bibnamefont {Temmink}},
  \bibinfo {author} {\bibfnamefont {M.}~\bibnamefont {Janssen}}, \bibinfo
  {author} {\bibfnamefont {C.}~\bibnamefont {Buisman}},\ and\ \bibinfo {author}
  {\bibfnamefont {R.}~\bibnamefont {Wijffels}},\ }\bibfield  {title} {\bibinfo
  {title} {Nitrogen and phosphorus removal from municipal wastewater effluent
  using microalgal biofilms},\ }\href@noop {} {\bibfield  {journal} {\bibinfo
  {journal} {Water Res.}\ }\textbf {\bibinfo {volume} {45}},\ \bibinfo {pages}
  {5925} (\bibinfo {year} {2011})}\BibitemShut {NoStop}%
\bibitem [{\citenamefont {Berner}\ \emph {et~al.}(2015)\citenamefont {Berner},
  \citenamefont {Heimann},\ and\ \citenamefont {Sheehan}}]{berner2015}%
  \BibitemOpen
  \bibfield  {author} {\bibinfo {author} {\bibfnamefont {F.}~\bibnamefont
  {Berner}}, \bibinfo {author} {\bibfnamefont {K.}~\bibnamefont {Heimann}},\
  and\ \bibinfo {author} {\bibfnamefont {M.}~\bibnamefont {Sheehan}},\
  }\bibfield  {title} {\bibinfo {title} {Microalgal biofilms for biomass
  production},\ }\href@noop {} {\bibfield  {journal} {\bibinfo  {journal} {J.
  Appl. Phycol.}\ }\textbf {\bibinfo {volume} {27}},\ \bibinfo {pages} {1793}
  (\bibinfo {year} {2015})}\BibitemShut {NoStop}%
\bibitem [{\citenamefont {Zeriouh}\ \emph {et~al.}(2017)\citenamefont
  {Zeriouh}, \citenamefont {Reinoso-Moreno}, \citenamefont {L{\'o}pez-Rosales},
  \citenamefont {Cer{\'o}n-Garc{\'\i}a}, \citenamefont {S{\'a}nchez-Mir{\'o}n},
  \citenamefont {Garc{\'\i}a-Camacho},\ and\ \citenamefont
  {Molina-Grima}}]{zeriouh2017biofouling}%
  \BibitemOpen
  \bibfield  {author} {\bibinfo {author} {\bibfnamefont {O.}~\bibnamefont
  {Zeriouh}}, \bibinfo {author} {\bibfnamefont {J.~V.}\ \bibnamefont
  {Reinoso-Moreno}}, \bibinfo {author} {\bibfnamefont {L.}~\bibnamefont
  {L{\'o}pez-Rosales}}, \bibinfo {author} {\bibfnamefont {M.~d.~C.}\
  \bibnamefont {Cer{\'o}n-Garc{\'\i}a}}, \bibinfo {author} {\bibfnamefont
  {A.}~\bibnamefont {S{\'a}nchez-Mir{\'o}n}}, \bibinfo {author} {\bibfnamefont
  {F.}~\bibnamefont {Garc{\'\i}a-Camacho}},\ and\ \bibinfo {author}
  {\bibfnamefont {E.}~\bibnamefont {Molina-Grima}},\ }\bibfield  {title}
  {\bibinfo {title} {Biofouling in photobioreactors for marine microalgae},\
  }\href@noop {} {\bibfield  {journal} {\bibinfo  {journal} {Critical reviews
  in biotechnology}\ }\textbf {\bibinfo {volume} {37}},\ \bibinfo {pages}
  {1006} (\bibinfo {year} {2017})}\BibitemShut {NoStop}%
\bibitem [{\citenamefont {Christenson}\ and\ \citenamefont
  {Sims}(2011)}]{christenson2011production}%
  \BibitemOpen
  \bibfield  {author} {\bibinfo {author} {\bibfnamefont {L.}~\bibnamefont
  {Christenson}}\ and\ \bibinfo {author} {\bibfnamefont {R.}~\bibnamefont
  {Sims}},\ }\bibfield  {title} {\bibinfo {title} {Production and harvesting of
  microalgae for wastewater treatment, biofuels, and bioproducts},\ }\href@noop
  {} {\bibfield  {journal} {\bibinfo  {journal} {Biotechnology advances}\
  }\textbf {\bibinfo {volume} {29}},\ \bibinfo {pages} {686} (\bibinfo {year}
  {2011})}\BibitemShut {NoStop}%
\bibitem [{\citenamefont {Roostaei}\ \emph {et~al.}(2018)\citenamefont
  {Roostaei}, \citenamefont {Zhang}, \citenamefont {Gopalakrishnan},\ and\
  \citenamefont {Ochocki}}]{roostaei2018mixotrophic}%
  \BibitemOpen
  \bibfield  {author} {\bibinfo {author} {\bibfnamefont {J.}~\bibnamefont
  {Roostaei}}, \bibinfo {author} {\bibfnamefont {Y.}~\bibnamefont {Zhang}},
  \bibinfo {author} {\bibfnamefont {K.}~\bibnamefont {Gopalakrishnan}},\ and\
  \bibinfo {author} {\bibfnamefont {A.~J.}\ \bibnamefont {Ochocki}},\
  }\bibfield  {title} {\bibinfo {title} {Mixotrophic microalgae biofilm: a
  novel algae cultivation strategy for improved productivity and
  cost-efficiency of biofuel feedstock production},\ }\href@noop {} {\bibfield
  {journal} {\bibinfo  {journal} {Sci. Rep.}\ }\textbf {\bibinfo {volume}
  {8}},\ \bibinfo {pages} {1} (\bibinfo {year} {2018})}\BibitemShut {NoStop}%
\bibitem [{\citenamefont {Schumann}\ \emph {et~al.}(2005)\citenamefont
  {Schumann}, \citenamefont {H{\"a}ubner}, \citenamefont {Klausch},\ and\
  \citenamefont {Karsten}}]{schumann2005chlorophyll}%
  \BibitemOpen
  \bibfield  {author} {\bibinfo {author} {\bibfnamefont {R.}~\bibnamefont
  {Schumann}}, \bibinfo {author} {\bibfnamefont {N.}~\bibnamefont
  {H{\"a}ubner}}, \bibinfo {author} {\bibfnamefont {S.}~\bibnamefont
  {Klausch}},\ and\ \bibinfo {author} {\bibfnamefont {U.}~\bibnamefont
  {Karsten}},\ }\bibfield  {title} {\bibinfo {title} {Chlorophyll extraction
  methods for the quantification of green microalgae colonizing building
  facades},\ }\href@noop {} {\bibfield  {journal} {\bibinfo  {journal} {Int.
  Biodeterior. Biodegradation}\ }\textbf {\bibinfo {volume} {55}},\ \bibinfo
  {pages} {213} (\bibinfo {year} {2005})}\BibitemShut {NoStop}%
\bibitem [{\citenamefont {Harris}(2001)}]{harris2001chlamydomonas}%
  \BibitemOpen
  \bibfield  {author} {\bibinfo {author} {\bibfnamefont {E.~H.}\ \bibnamefont
  {Harris}},\ }\bibfield  {title} {\bibinfo {title} {Chlamydomonas as a model
  organism},\ }\href@noop {} {\bibfield  {journal} {\bibinfo  {journal} {Annu.
  Rev. Plant Biol.}\ }\textbf {\bibinfo {volume} {52}},\ \bibinfo {pages} {363}
  (\bibinfo {year} {2001})}\BibitemShut {NoStop}%
\bibitem [{\citenamefont {Hippler}\ \emph {et~al.}(1998)\citenamefont
  {Hippler}, \citenamefont {Redding},\ and\ \citenamefont
  {Rochaix}}]{hippler1998chlamydomonas}%
  \BibitemOpen
  \bibfield  {author} {\bibinfo {author} {\bibfnamefont {M.}~\bibnamefont
  {Hippler}}, \bibinfo {author} {\bibfnamefont {K.}~\bibnamefont {Redding}},\
  and\ \bibinfo {author} {\bibfnamefont {J.-D.}\ \bibnamefont {Rochaix}},\
  }\bibfield  {title} {\bibinfo {title} {Chlamydomonas genetics, a tool for the
  study of bioenergetic pathways},\ }\href@noop {} {\bibfield  {journal}
  {\bibinfo  {journal} {Biochimica et Biophysica Acta-Bioenergetics}\ }\textbf
  {\bibinfo {volume} {1367}},\ \bibinfo {pages} {1} (\bibinfo {year}
  {1998})}\BibitemShut {NoStop}%
\bibitem [{\citenamefont {Rochaix}\ \emph {et~al.}(2006)\citenamefont
  {Rochaix}, \citenamefont {Goldschmidt-Clermont},\ and\ \citenamefont
  {Merchant}}]{rochaix2006molecular}%
  \BibitemOpen
  \bibfield  {author} {\bibinfo {author} {\bibfnamefont {J.-D.}\ \bibnamefont
  {Rochaix}}, \bibinfo {author} {\bibfnamefont {M.}~\bibnamefont
  {Goldschmidt-Clermont}},\ and\ \bibinfo {author} {\bibfnamefont
  {S.}~\bibnamefont {Merchant}},\ }\href@noop {} {\emph {\bibinfo {title} {The
  molecular biology of chloroplasts and mitochondria in Chlamydomonas}}},\
  Vol.~\bibinfo {volume} {7}\ (\bibinfo  {publisher} {Springer Science \&
  Business Media},\ \bibinfo {year} {2006})\BibitemShut {NoStop}%
\bibitem [{\citenamefont {Salom{\'e}}\ and\ \citenamefont
  {Merchant}(2019)}]{salome2019series}%
  \BibitemOpen
  \bibfield  {author} {\bibinfo {author} {\bibfnamefont {P.~A.}\ \bibnamefont
  {Salom{\'e}}}\ and\ \bibinfo {author} {\bibfnamefont {S.~S.}\ \bibnamefont
  {Merchant}},\ }\bibfield  {title} {\bibinfo {title} {A series of fortunate
  events: Introducing chlamydomonas as a reference organism},\ }\href@noop {}
  {\bibfield  {journal} {\bibinfo  {journal} {Plant Cell}\ }\textbf {\bibinfo
  {volume} {31}},\ \bibinfo {pages} {1682} (\bibinfo {year}
  {2019})}\BibitemShut {NoStop}%
\bibitem [{\citenamefont {Goldstein}\ \emph {et~al.}(2011)\citenamefont
  {Goldstein}, \citenamefont {Polin},\ and\ \citenamefont
  {Tuval}}]{goldstein2011}%
  \BibitemOpen
  \bibfield  {author} {\bibinfo {author} {\bibfnamefont {R.~E.}\ \bibnamefont
  {Goldstein}}, \bibinfo {author} {\bibfnamefont {M.}~\bibnamefont {Polin}},\
  and\ \bibinfo {author} {\bibfnamefont {I.}~\bibnamefont {Tuval}},\ }\bibfield
   {title} {\bibinfo {title} {Emergence of synchronized beating during the
  regrowth of eukaryotic flagella},\ }\href
  {https://doi.org/10.1103/PhysRevLett.107.148103} {\bibfield  {journal}
  {\bibinfo  {journal} {Phys. Rev. Lett.}\ }\textbf {\bibinfo {volume} {107}},\
  \bibinfo {pages} {148103} (\bibinfo {year} {2011})}\BibitemShut {NoStop}%
\bibitem [{\citenamefont {Geyer}\ \emph {et~al.}(2013)\citenamefont {Geyer},
  \citenamefont {Jülicher}, \citenamefont {Howard},\ and\ \citenamefont
  {Friedrich}}]{geyer2013}%
  \BibitemOpen
  \bibfield  {author} {\bibinfo {author} {\bibfnamefont {V.~F.}\ \bibnamefont
  {Geyer}}, \bibinfo {author} {\bibfnamefont {F.}~\bibnamefont {Jülicher}},
  \bibinfo {author} {\bibfnamefont {J.}~\bibnamefont {Howard}},\ and\ \bibinfo
  {author} {\bibfnamefont {B.~M.}\ \bibnamefont {Friedrich}},\ }\bibfield
  {title} {\bibinfo {title} {Cell-body rocking is a dominant mechanism for
  flagellar synchronization in a swimming alga},\ }\href
  {https://doi.org/10.1073/pnas.1300895110} {\bibfield  {journal} {\bibinfo
  {journal} {Proceedings of the National Academy of Sciences}\ }\textbf
  {\bibinfo {volume} {110}},\ \bibinfo {pages} {18058} (\bibinfo {year}
  {2013})},\ \Eprint
  {https://arxiv.org/abs/https://www.pnas.org/doi/pdf/10.1073/pnas.1300895110}
  {https://www.pnas.org/doi/pdf/10.1073/pnas.1300895110} \BibitemShut {NoStop}%
\bibitem [{\citenamefont {Wan}\ and\ \citenamefont
  {Goldstein}(2016)}]{wan2016}%
  \BibitemOpen
  \bibfield  {author} {\bibinfo {author} {\bibfnamefont {K.~Y.}\ \bibnamefont
  {Wan}}\ and\ \bibinfo {author} {\bibfnamefont {R.~E.}\ \bibnamefont
  {Goldstein}},\ }\bibfield  {title} {\bibinfo {title} {Coordinated beating of
  algal flagella is mediated by basal coupling},\ }\href
  {https://doi.org/10.1073/pnas.1518527113} {\bibfield  {journal} {\bibinfo
  {journal} {Proceedings of the National Academy of Sciences}\ }\textbf
  {\bibinfo {volume} {113}},\ \bibinfo {pages} {E2784} (\bibinfo {year}
  {2016})},\ \Eprint
  {https://arxiv.org/abs/https://www.pnas.org/doi/pdf/10.1073/pnas.1518527113}
  {https://www.pnas.org/doi/pdf/10.1073/pnas.1518527113} \BibitemShut {NoStop}%
\bibitem [{\citenamefont {Böddeker}\ \emph {et~al.}(2020)\citenamefont
  {Böddeker}, \citenamefont {Karpitschka}, \citenamefont {Kreis},
  \citenamefont {Magdelaine},\ and\ \citenamefont {Bäumchen}}]{boeddeker2020}%
  \BibitemOpen
  \bibfield  {author} {\bibinfo {author} {\bibfnamefont {T.~J.}\ \bibnamefont
  {Böddeker}}, \bibinfo {author} {\bibfnamefont {S.}~\bibnamefont
  {Karpitschka}}, \bibinfo {author} {\bibfnamefont {C.~T.}\ \bibnamefont
  {Kreis}}, \bibinfo {author} {\bibfnamefont {Q.}~\bibnamefont {Magdelaine}},\
  and\ \bibinfo {author} {\bibfnamefont {O.}~\bibnamefont {Bäumchen}},\
  }\bibfield  {title} {\bibinfo {title} {Dynamic force measurements on swimming
  <i>chlamydomonas</i> cells using micropipette force sensors},\ }\href
  {https://doi.org/10.1098/rsif.2019.0580} {\bibfield  {journal} {\bibinfo
  {journal} {Journal of The Royal Society Interface}\ }\textbf {\bibinfo
  {volume} {17}},\ \bibinfo {pages} {20190580} (\bibinfo {year} {2020})},\
  \Eprint
  {https://arxiv.org/abs/https://royalsocietypublishing.org/doi/pdf/10.1098/rsif.2019.0580}
  {https://royalsocietypublishing.org/doi/pdf/10.1098/rsif.2019.0580}
  \BibitemShut {NoStop}%
\bibitem [{\citenamefont {Polin}\ \emph {et~al.}(2009)\citenamefont {Polin},
  \citenamefont {Tuval}, \citenamefont {Drescher}, \citenamefont {Gollub},\
  and\ \citenamefont {Goldstein}}]{polin2009}%
  \BibitemOpen
  \bibfield  {author} {\bibinfo {author} {\bibfnamefont {M.}~\bibnamefont
  {Polin}}, \bibinfo {author} {\bibfnamefont {I.}~\bibnamefont {Tuval}},
  \bibinfo {author} {\bibfnamefont {K.}~\bibnamefont {Drescher}}, \bibinfo
  {author} {\bibfnamefont {J.~P.}\ \bibnamefont {Gollub}},\ and\ \bibinfo
  {author} {\bibfnamefont {R.~E.}\ \bibnamefont {Goldstein}},\ }\bibfield
  {title} {\bibinfo {title} {<i>chlamydomonas</i> swims with two
  \&\#x201c;gears\&\#x201d; in a eukaryotic version of run-and-tumble
  locomotion},\ }\href {https://doi.org/10.1126/science.1172667} {\bibfield
  {journal} {\bibinfo  {journal} {Science}\ }\textbf {\bibinfo {volume}
  {325}},\ \bibinfo {pages} {487} (\bibinfo {year} {2009})},\ \Eprint
  {https://arxiv.org/abs/https://www.science.org/doi/pdf/10.1126/science.1172667}
  {https://www.science.org/doi/pdf/10.1126/science.1172667} \BibitemShut
  {NoStop}%
\bibitem [{\citenamefont {Ostapenko}\ \emph {et~al.}(2018)\citenamefont
  {Ostapenko}, \citenamefont {Schwarzendahl}, \citenamefont {B\"oddeker},
  \citenamefont {Kreis}, \citenamefont {Cammann}, \citenamefont {Mazza},\ and\
  \citenamefont {B\"aumchen}}]{ostapenko2018}%
  \BibitemOpen
  \bibfield  {author} {\bibinfo {author} {\bibfnamefont {T.}~\bibnamefont
  {Ostapenko}}, \bibinfo {author} {\bibfnamefont {F.~J.}\ \bibnamefont
  {Schwarzendahl}}, \bibinfo {author} {\bibfnamefont {T.~J.}\ \bibnamefont
  {B\"oddeker}}, \bibinfo {author} {\bibfnamefont {C.~T.}\ \bibnamefont
  {Kreis}}, \bibinfo {author} {\bibfnamefont {J.}~\bibnamefont {Cammann}},
  \bibinfo {author} {\bibfnamefont {M.~G.}\ \bibnamefont {Mazza}},\ and\
  \bibinfo {author} {\bibfnamefont {O.}~\bibnamefont {B\"aumchen}},\ }\bibfield
   {title} {\bibinfo {title} {Curvature-guided motility of microalgae in
  geometric confinement},\ }\href
  {https://doi.org/10.1103/PhysRevLett.120.068002} {\bibfield  {journal}
  {\bibinfo  {journal} {Phys. Rev. Lett.}\ }\textbf {\bibinfo {volume} {120}},\
  \bibinfo {pages} {068002} (\bibinfo {year} {2018})}\BibitemShut {NoStop}%
\bibitem [{\citenamefont {Cammann}\ \emph {et~al.}(2021)\citenamefont
  {Cammann}, \citenamefont {Schwarzendahl}, \citenamefont {Ostapenko},
  \citenamefont {Lavrentovich}, \citenamefont {Bäumchen},\ and\ \citenamefont
  {Mazza}}]{cammann2021}%
  \BibitemOpen
  \bibfield  {author} {\bibinfo {author} {\bibfnamefont {J.}~\bibnamefont
  {Cammann}}, \bibinfo {author} {\bibfnamefont {F.~J.}\ \bibnamefont
  {Schwarzendahl}}, \bibinfo {author} {\bibfnamefont {T.}~\bibnamefont
  {Ostapenko}}, \bibinfo {author} {\bibfnamefont {D.}~\bibnamefont
  {Lavrentovich}}, \bibinfo {author} {\bibfnamefont {O.}~\bibnamefont
  {Bäumchen}},\ and\ \bibinfo {author} {\bibfnamefont {M.~G.}\ \bibnamefont
  {Mazza}},\ }\bibfield  {title} {\bibinfo {title} {Emergent probability fluxes
  in confined microbial navigation},\ }\href
  {https://doi.org/10.1073/pnas.2024752118} {\bibfield  {journal} {\bibinfo
  {journal} {Proceedings of the National Academy of Sciences}\ }\textbf
  {\bibinfo {volume} {118}},\ \bibinfo {pages} {e2024752118} (\bibinfo {year}
  {2021})},\ \Eprint
  {https://arxiv.org/abs/https://www.pnas.org/doi/pdf/10.1073/pnas.2024752118}
  {https://www.pnas.org/doi/pdf/10.1073/pnas.2024752118} \BibitemShut {NoStop}%
\bibitem [{\citenamefont {Bloodgood}\ and\ \citenamefont
  {Workman}(1984)}]{bloodgood1984flagellar}%
  \BibitemOpen
  \bibfield  {author} {\bibinfo {author} {\bibfnamefont {R.~A.}\ \bibnamefont
  {Bloodgood}}\ and\ \bibinfo {author} {\bibfnamefont {L.~J.}\ \bibnamefont
  {Workman}},\ }\bibfield  {title} {\bibinfo {title} {A flagellar surface
  glycoprotein mediating cell-substrate interaction in chlamydomonas},\
  }\href@noop {} {\bibfield  {journal} {\bibinfo  {journal} {Cell motility}\
  }\textbf {\bibinfo {volume} {4}},\ \bibinfo {pages} {77} (\bibinfo {year}
  {1984})}\BibitemShut {NoStop}%
\bibitem [{\citenamefont {Bloodgood}\ \emph {et~al.}(2019)\citenamefont
  {Bloodgood}, \citenamefont {Tetreault},\ and\ \citenamefont
  {Sloboda}}]{bloodgood2019chlamydomonas}%
  \BibitemOpen
  \bibfield  {author} {\bibinfo {author} {\bibfnamefont {R.~A.}\ \bibnamefont
  {Bloodgood}}, \bibinfo {author} {\bibfnamefont {J.}~\bibnamefont
  {Tetreault}},\ and\ \bibinfo {author} {\bibfnamefont {R.~D.}\ \bibnamefont
  {Sloboda}},\ }\bibfield  {title} {\bibinfo {title} {The chlamydomonas
  flagellar membrane glycoprotein fmg-1b is necessary for expression of force
  at the flagellar surface},\ }\href@noop {} {\bibfield  {journal} {\bibinfo
  {journal} {Journal of cell science}\ }\textbf {\bibinfo {volume} {132}},\
  \bibinfo {pages} {jcs233429} (\bibinfo {year} {2019})}\BibitemShut {NoStop}%
\bibitem [{\citenamefont {Bloodgood}(1981)}]{bloodgood1981gliding}%
  \BibitemOpen
  \bibfield  {author} {\bibinfo {author} {\bibfnamefont {R.}~\bibnamefont
  {Bloodgood}},\ }\bibfield  {title} {\bibinfo {title} {Flagella-dependent
  gliding motility in chlamydomonas},\ }\href@noop {} {\bibfield  {journal}
  {\bibinfo  {journal} {Protoplasma}\ }\textbf {\bibinfo {volume} {106}},\
  \bibinfo {pages} {183} (\bibinfo {year} {1981})}\BibitemShut {NoStop}%
\bibitem [{\citenamefont {Kreis}\ \emph {et~al.}(2018)\citenamefont {Kreis},
  \citenamefont {Le~Blay}, \citenamefont {Linne}, \citenamefont {Makowski},\
  and\ \citenamefont {B{\"a}umchen}}]{kreis2018adhesion}%
  \BibitemOpen
  \bibfield  {author} {\bibinfo {author} {\bibfnamefont {C.~T.}\ \bibnamefont
  {Kreis}}, \bibinfo {author} {\bibfnamefont {M.}~\bibnamefont {Le~Blay}},
  \bibinfo {author} {\bibfnamefont {C.}~\bibnamefont {Linne}}, \bibinfo
  {author} {\bibfnamefont {M.~M.}\ \bibnamefont {Makowski}},\ and\ \bibinfo
  {author} {\bibfnamefont {O.}~\bibnamefont {B{\"a}umchen}},\ }\bibfield
  {title} {\bibinfo {title} {Adhesion of chlamydomonas microalgae to surfaces
  is switchable by light},\ }\href@noop {} {\bibfield  {journal} {\bibinfo
  {journal} {Nat. Phys.}\ }\textbf {\bibinfo {volume} {14}},\ \bibinfo {pages}
  {45} (\bibinfo {year} {2018})}\BibitemShut {NoStop}%
\bibitem [{\citenamefont {Kreis}\ \emph {et~al.}(2019)\citenamefont {Kreis},
  \citenamefont {Grangier},\ and\ \citenamefont
  {B{\"a}umchen}}]{kreis2019vivo}%
  \BibitemOpen
  \bibfield  {author} {\bibinfo {author} {\bibfnamefont {C.~T.}\ \bibnamefont
  {Kreis}}, \bibinfo {author} {\bibfnamefont {A.}~\bibnamefont {Grangier}},\
  and\ \bibinfo {author} {\bibfnamefont {O.}~\bibnamefont {B{\"a}umchen}},\
  }\bibfield  {title} {\bibinfo {title} {In vivo adhesion force measurements of
  chlamydomonas on model substrates},\ }\href@noop {} {\bibfield  {journal}
  {\bibinfo  {journal} {Soft Matter}\ }\textbf {\bibinfo {volume} {15}},\
  \bibinfo {pages} {3027} (\bibinfo {year} {2019})}\BibitemShut {NoStop}%
\bibitem [{\citenamefont {Xu}\ \emph {et~al.}(2020)\citenamefont {Xu},
  \citenamefont {Oltmanns}, \citenamefont {Zhao}, \citenamefont {Girot},
  \citenamefont {Karimi}, \citenamefont {Hoepfner}, \citenamefont {Kelterborn},
  \citenamefont {Scholz}, \citenamefont {Beißel}, \citenamefont {Hegemann},
  \citenamefont {Bäumchen}, \citenamefont {Liu}, \citenamefont {Huang},\ and\
  \citenamefont {Hippler}}]{xu2020}%
  \BibitemOpen
  \bibfield  {author} {\bibinfo {author} {\bibfnamefont {N.}~\bibnamefont
  {Xu}}, \bibinfo {author} {\bibfnamefont {A.}~\bibnamefont {Oltmanns}},
  \bibinfo {author} {\bibfnamefont {L.}~\bibnamefont {Zhao}}, \bibinfo {author}
  {\bibfnamefont {A.}~\bibnamefont {Girot}}, \bibinfo {author} {\bibfnamefont
  {M.}~\bibnamefont {Karimi}}, \bibinfo {author} {\bibfnamefont
  {L.}~\bibnamefont {Hoepfner}}, \bibinfo {author} {\bibfnamefont
  {S.}~\bibnamefont {Kelterborn}}, \bibinfo {author} {\bibfnamefont
  {M.}~\bibnamefont {Scholz}}, \bibinfo {author} {\bibfnamefont
  {J.}~\bibnamefont {Beißel}}, \bibinfo {author} {\bibfnamefont
  {P.}~\bibnamefont {Hegemann}}, \bibinfo {author} {\bibfnamefont
  {O.}~\bibnamefont {Bäumchen}}, \bibinfo {author} {\bibfnamefont {L.-N.}\
  \bibnamefont {Liu}}, \bibinfo {author} {\bibfnamefont {K.}~\bibnamefont
  {Huang}},\ and\ \bibinfo {author} {\bibfnamefont {M.}~\bibnamefont
  {Hippler}},\ }\bibfield  {title} {\bibinfo {title} {Altered \textit{N}-glycan
  composition impacts flagella-mediated adhesion in \textit{Chlamydomonas
  reinhardtii}},\ }\href {https://doi.org/10.7554/eLife.58805} {\bibfield
  {journal} {\bibinfo  {journal} {eLife}\ }\textbf {\bibinfo {volume} {9}},\
  \bibinfo {pages} {e58805} (\bibinfo {year} {2020})}\BibitemShut {NoStop}%
\bibitem [{\citenamefont {Stavis}\ and\ \citenamefont
  {Hirschberg}(1973)}]{stavis1973phototaxis}%
  \BibitemOpen
  \bibfield  {author} {\bibinfo {author} {\bibfnamefont {R.~L.}\ \bibnamefont
  {Stavis}}\ and\ \bibinfo {author} {\bibfnamefont {R.}~\bibnamefont
  {Hirschberg}},\ }\bibfield  {title} {\bibinfo {title} {Phototaxis in
  chlamydomonas reinhardtii},\ }\href@noop {} {\bibfield  {journal} {\bibinfo
  {journal} {The Journal of cell biology}\ }\textbf {\bibinfo {volume} {59}},\
  \bibinfo {pages} {367} (\bibinfo {year} {1973})}\BibitemShut {NoStop}%
\bibitem [{\citenamefont {BEAN}(1977)}]{bean1977geotactic}%
  \BibitemOpen
  \bibfield  {author} {\bibinfo {author} {\bibfnamefont {B.}~\bibnamefont
  {BEAN}},\ }\bibfield  {title} {\bibinfo {title} {Geotactic behavior of
  chlamydomonas},\ }\href@noop {} {\bibfield  {journal} {\bibinfo  {journal}
  {The Journal of Protozoology}\ }\textbf {\bibinfo {volume} {24}},\ \bibinfo
  {pages} {394} (\bibinfo {year} {1977})}\BibitemShut {NoStop}%
\bibitem [{\citenamefont {Randall}\ \emph {et~al.}(1968)\citenamefont
  {Randall}, \citenamefont {Cavalier-Smith}, \citenamefont {MCVITTIE},
  \citenamefont {Warr},\ and\ \citenamefont
  {Hopkins}}]{randall1968developmental}%
  \BibitemOpen
  \bibfield  {author} {\bibinfo {author} {\bibfnamefont {J.}~\bibnamefont
  {Randall}}, \bibinfo {author} {\bibfnamefont {T.}~\bibnamefont
  {Cavalier-Smith}}, \bibinfo {author} {\bibfnamefont {A.}~\bibnamefont
  {MCVITTIE}}, \bibinfo {author} {\bibfnamefont {J.}~\bibnamefont {Warr}},\
  and\ \bibinfo {author} {\bibfnamefont {J.}~\bibnamefont {Hopkins}},\
  }\bibfield  {title} {\bibinfo {title} {Developmental and control processes in
  the basal bodies and flagella of chlamydomonas reinhardii},\ }in\ \href@noop
  {} {\emph {\bibinfo {booktitle} {Dev. Biol.}}}\ (\bibinfo  {publisher}
  {Elsevier},\ \bibinfo {year} {1968})\ pp.\ \bibinfo {pages}
  {43--83}\BibitemShut {NoStop}%
\bibitem [{\citenamefont {Rosenbaum}\ \emph {et~al.}(1969)\citenamefont
  {Rosenbaum}, \citenamefont {Moulder},\ and\ \citenamefont
  {Ringo}}]{rosenbaum1969flagellar}%
  \BibitemOpen
  \bibfield  {author} {\bibinfo {author} {\bibfnamefont {J.~L.}\ \bibnamefont
  {Rosenbaum}}, \bibinfo {author} {\bibfnamefont {J.~E.}\ \bibnamefont
  {Moulder}},\ and\ \bibinfo {author} {\bibfnamefont {D.~L.}\ \bibnamefont
  {Ringo}},\ }\bibfield  {title} {\bibinfo {title} {Flagellar elongation and
  shortening in chlamydomonas: the use of cycloheximide and colchicine to study
  the synthesis and assembly of flagellar proteins},\ }\href@noop {} {\bibfield
   {journal} {\bibinfo  {journal} {J. Cell Biol.}\ }\textbf {\bibinfo {volume}
  {41}},\ \bibinfo {pages} {600} (\bibinfo {year} {1969})}\BibitemShut
  {NoStop}%
\bibitem [{\citenamefont {Kam}\ \emph {et~al.}(1999)\citenamefont {Kam},
  \citenamefont {Moseyko}, \citenamefont {Nemson},\ and\ \citenamefont
  {Feldman}}]{kam1999gravitaxis}%
  \BibitemOpen
  \bibfield  {author} {\bibinfo {author} {\bibfnamefont {V.}~\bibnamefont
  {Kam}}, \bibinfo {author} {\bibfnamefont {N.}~\bibnamefont {Moseyko}},
  \bibinfo {author} {\bibfnamefont {J.}~\bibnamefont {Nemson}},\ and\ \bibinfo
  {author} {\bibfnamefont {L.~J.}\ \bibnamefont {Feldman}},\ }\bibfield
  {title} {\bibinfo {title} {Gravitaxis in chlamydomonas reinhardtii:
  characterization using video microscopy and computer analysis},\ }\href@noop
  {} {\bibfield  {journal} {\bibinfo  {journal} {Int. J. Plant Sci.}\ }\textbf
  {\bibinfo {volume} {160}},\ \bibinfo {pages} {1093} (\bibinfo {year}
  {1999})}\BibitemShut {NoStop}%
\bibitem [{\citenamefont {Arrieta}\ \emph {et~al.}(2017)\citenamefont
  {Arrieta}, \citenamefont {Barreira}, \citenamefont {Chioccioli},
  \citenamefont {Polin},\ and\ \citenamefont {Tuval}}]{arrieta2017phototaxis}%
  \BibitemOpen
  \bibfield  {author} {\bibinfo {author} {\bibfnamefont {J.}~\bibnamefont
  {Arrieta}}, \bibinfo {author} {\bibfnamefont {A.}~\bibnamefont {Barreira}},
  \bibinfo {author} {\bibfnamefont {M.}~\bibnamefont {Chioccioli}}, \bibinfo
  {author} {\bibfnamefont {M.}~\bibnamefont {Polin}},\ and\ \bibinfo {author}
  {\bibfnamefont {I.}~\bibnamefont {Tuval}},\ }\bibfield  {title} {\bibinfo
  {title} {Phototaxis beyond turning: persistent accumulation and response
  acclimation of the microalga chlamydomonas reinhardtii},\ }\href@noop {}
  {\bibfield  {journal} {\bibinfo  {journal} {Scientific reports}\ }\textbf
  {\bibinfo {volume} {7}},\ \bibinfo {pages} {1} (\bibinfo {year}
  {2017})}\BibitemShut {NoStop}%
\bibitem [{\citenamefont {Greiner}\ \emph {et~al.}(2017)\citenamefont
  {Greiner}, \citenamefont {Kelterborn}, \citenamefont {Evers}, \citenamefont
  {Kreimer}, \citenamefont {Sizova},\ and\ \citenamefont
  {Hegemann}}]{greiner2017targeting}%
  \BibitemOpen
  \bibfield  {author} {\bibinfo {author} {\bibfnamefont {A.}~\bibnamefont
  {Greiner}}, \bibinfo {author} {\bibfnamefont {S.}~\bibnamefont {Kelterborn}},
  \bibinfo {author} {\bibfnamefont {H.}~\bibnamefont {Evers}}, \bibinfo
  {author} {\bibfnamefont {G.}~\bibnamefont {Kreimer}}, \bibinfo {author}
  {\bibfnamefont {I.}~\bibnamefont {Sizova}},\ and\ \bibinfo {author}
  {\bibfnamefont {P.}~\bibnamefont {Hegemann}},\ }\bibfield  {title} {\bibinfo
  {title} {Targeting of photoreceptor genes in chlamydomonas reinhardtii via
  zinc-finger nucleases and crispr/cas9},\ }\href@noop {} {\bibfield  {journal}
  {\bibinfo  {journal} {The Plant Cell}\ }\textbf {\bibinfo {volume} {29}},\
  \bibinfo {pages} {2498} (\bibinfo {year} {2017})}\BibitemShut {NoStop}%
\bibitem [{\citenamefont {Nagel}\ \emph {et~al.}(2002)\citenamefont {Nagel},
  \citenamefont {Ollig}, \citenamefont {Fuhrmann}, \citenamefont {Kateriya},
  \citenamefont {Musti}, \citenamefont {Bamberg},\ and\ \citenamefont
  {Hegemann}}]{nagel2002channelrhodopsin}%
  \BibitemOpen
  \bibfield  {author} {\bibinfo {author} {\bibfnamefont {G.}~\bibnamefont
  {Nagel}}, \bibinfo {author} {\bibfnamefont {D.}~\bibnamefont {Ollig}},
  \bibinfo {author} {\bibfnamefont {M.}~\bibnamefont {Fuhrmann}}, \bibinfo
  {author} {\bibfnamefont {S.}~\bibnamefont {Kateriya}}, \bibinfo {author}
  {\bibfnamefont {A.~M.}\ \bibnamefont {Musti}}, \bibinfo {author}
  {\bibfnamefont {E.}~\bibnamefont {Bamberg}},\ and\ \bibinfo {author}
  {\bibfnamefont {P.}~\bibnamefont {Hegemann}},\ }\bibfield  {title} {\bibinfo
  {title} {Channelrhodopsin-1: a light-gated proton channel in green algae},\
  }\href@noop {} {\bibfield  {journal} {\bibinfo  {journal} {Science}\ }\textbf
  {\bibinfo {volume} {296}},\ \bibinfo {pages} {2395} (\bibinfo {year}
  {2002})}\BibitemShut {NoStop}%
\bibitem [{\citenamefont {Nagel}\ \emph {et~al.}(2003)\citenamefont {Nagel},
  \citenamefont {Szellas}, \citenamefont {Huhn}, \citenamefont {Kateriya},
  \citenamefont {Adeishvili}, \citenamefont {Berthold}, \citenamefont {Ollig},
  \citenamefont {Hegemann},\ and\ \citenamefont
  {Bamberg}}]{nagel2003channelrhodopsin}%
  \BibitemOpen
  \bibfield  {author} {\bibinfo {author} {\bibfnamefont {G.}~\bibnamefont
  {Nagel}}, \bibinfo {author} {\bibfnamefont {T.}~\bibnamefont {Szellas}},
  \bibinfo {author} {\bibfnamefont {W.}~\bibnamefont {Huhn}}, \bibinfo {author}
  {\bibfnamefont {S.}~\bibnamefont {Kateriya}}, \bibinfo {author}
  {\bibfnamefont {N.}~\bibnamefont {Adeishvili}}, \bibinfo {author}
  {\bibfnamefont {P.}~\bibnamefont {Berthold}}, \bibinfo {author}
  {\bibfnamefont {D.}~\bibnamefont {Ollig}}, \bibinfo {author} {\bibfnamefont
  {P.}~\bibnamefont {Hegemann}},\ and\ \bibinfo {author} {\bibfnamefont
  {E.}~\bibnamefont {Bamberg}},\ }\bibfield  {title} {\bibinfo {title}
  {Channelrhodopsin-2, a directly light-gated cation-selective membrane
  channel},\ }\href@noop {} {\bibfield  {journal} {\bibinfo  {journal} {Proc.
  Natl. Acad. Sci. U. S. A.}\ }\textbf {\bibinfo {volume} {100}},\ \bibinfo
  {pages} {13940} (\bibinfo {year} {2003})}\BibitemShut {NoStop}%
\bibitem [{\citenamefont {Chenouard}\ \emph {et~al.}(2014)\citenamefont
  {Chenouard}, \citenamefont {Smal}, \citenamefont {De~Chaumont}, \citenamefont
  {Ma{\v{s}}ka}, \citenamefont {Sbalzarini}, \citenamefont {Gong},
  \citenamefont {Cardinale}, \citenamefont {Carthel}, \citenamefont
  {Coraluppi}, \citenamefont {Winter} \emph {et~al.}}]{chenouard2014objective}%
  \BibitemOpen
  \bibfield  {author} {\bibinfo {author} {\bibfnamefont {N.}~\bibnamefont
  {Chenouard}}, \bibinfo {author} {\bibfnamefont {I.}~\bibnamefont {Smal}},
  \bibinfo {author} {\bibfnamefont {F.}~\bibnamefont {De~Chaumont}}, \bibinfo
  {author} {\bibfnamefont {M.}~\bibnamefont {Ma{\v{s}}ka}}, \bibinfo {author}
  {\bibfnamefont {I.~F.}\ \bibnamefont {Sbalzarini}}, \bibinfo {author}
  {\bibfnamefont {Y.}~\bibnamefont {Gong}}, \bibinfo {author} {\bibfnamefont
  {J.}~\bibnamefont {Cardinale}}, \bibinfo {author} {\bibfnamefont
  {C.}~\bibnamefont {Carthel}}, \bibinfo {author} {\bibfnamefont
  {S.}~\bibnamefont {Coraluppi}}, \bibinfo {author} {\bibfnamefont
  {M.}~\bibnamefont {Winter}}, \emph {et~al.},\ }\bibfield  {title} {\bibinfo
  {title} {Objective comparison of particle tracking methods},\ }\href@noop {}
  {\bibfield  {journal} {\bibinfo  {journal} {Nat. Methods}\ }\textbf {\bibinfo
  {volume} {11}},\ \bibinfo {pages} {281} (\bibinfo {year} {2014})}\BibitemShut
  {NoStop}%
\bibitem [{\citenamefont {Meijering}\ \emph {et~al.}(2012)\citenamefont
  {Meijering}, \citenamefont {Dzyubachyk},\ and\ \citenamefont
  {Smal}}]{meijering2012methods}%
  \BibitemOpen
  \bibfield  {author} {\bibinfo {author} {\bibfnamefont {E.}~\bibnamefont
  {Meijering}}, \bibinfo {author} {\bibfnamefont {O.}~\bibnamefont
  {Dzyubachyk}},\ and\ \bibinfo {author} {\bibfnamefont {I.}~\bibnamefont
  {Smal}},\ }\bibfield  {title} {\bibinfo {title} {Methods for cell and
  particle tracking},\ }in\ \href@noop {} {\emph {\bibinfo {booktitle} {Methods
  Enzymol.}}},\ Vol.\ \bibinfo {volume} {504}\ (\bibinfo  {publisher}
  {Elsevier},\ \bibinfo {year} {2012})\ pp.\ \bibinfo {pages}
  {183--200}\BibitemShut {NoStop}%
\bibitem [{\citenamefont {Duda}\ and\ \citenamefont
  {Hart}(1972)}]{duda1972use}%
  \BibitemOpen
  \bibfield  {author} {\bibinfo {author} {\bibfnamefont {R.~O.}\ \bibnamefont
  {Duda}}\ and\ \bibinfo {author} {\bibfnamefont {P.~E.}\ \bibnamefont
  {Hart}},\ }\bibfield  {title} {\bibinfo {title} {Use of the hough
  transformation to detect lines and curves in pictures},\ }\href@noop {}
  {\bibfield  {journal} {\bibinfo  {journal} {Commun. ACM}\ }\textbf {\bibinfo
  {volume} {15}},\ \bibinfo {pages} {11} (\bibinfo {year} {1972})}\BibitemShut
  {NoStop}%
\bibitem [{\citenamefont {Blair}\ and\ \citenamefont
  {Dufresne}(2008)}]{blair2008matlab}%
  \BibitemOpen
  \bibfield  {author} {\bibinfo {author} {\bibfnamefont {D.}~\bibnamefont
  {Blair}}\ and\ \bibinfo {author} {\bibfnamefont {E.}~\bibnamefont
  {Dufresne}},\ }\bibfield  {title} {\bibinfo {title} {The matlab particle
  tracking code repository},\ }\href@noop {} {\bibfield  {journal} {\bibinfo
  {journal} {Particle-tracking code available at
  http://physics.georgetown.edu/matlab}\ } (\bibinfo {year}
  {2008})}\BibitemShut {NoStop}%
\bibitem [{\citenamefont {Shih}\ \emph {et~al.}(2013)\citenamefont {Shih},
  \citenamefont {Engel}, \citenamefont {Kocabas}, \citenamefont {Bilyard},
  \citenamefont {Gennerich}, \citenamefont {Marshall},\ and\ \citenamefont
  {Yildiz}}]{shih2013intraflagellar}%
  \BibitemOpen
  \bibfield  {author} {\bibinfo {author} {\bibfnamefont {S.~M.}\ \bibnamefont
  {Shih}}, \bibinfo {author} {\bibfnamefont {B.~D.}\ \bibnamefont {Engel}},
  \bibinfo {author} {\bibfnamefont {F.}~\bibnamefont {Kocabas}}, \bibinfo
  {author} {\bibfnamefont {T.}~\bibnamefont {Bilyard}}, \bibinfo {author}
  {\bibfnamefont {A.}~\bibnamefont {Gennerich}}, \bibinfo {author}
  {\bibfnamefont {W.~F.}\ \bibnamefont {Marshall}},\ and\ \bibinfo {author}
  {\bibfnamefont {A.}~\bibnamefont {Yildiz}},\ }\bibfield  {title} {\bibinfo
  {title} {Intraflagellar transport drives flagellar surface motility},\
  }\href@noop {} {\bibfield  {journal} {\bibinfo  {journal} {Elife}\ }\textbf
  {\bibinfo {volume} {2}},\ \bibinfo {pages} {e00744} (\bibinfo {year}
  {2013})}\BibitemShut {NoStop}%
\bibitem [{\citenamefont {Langmuir}(1918)}]{langmuir1918adsorption}%
  \BibitemOpen
  \bibfield  {author} {\bibinfo {author} {\bibfnamefont {I.}~\bibnamefont
  {Langmuir}},\ }\bibfield  {title} {\bibinfo {title} {The adsorption of gases
  on plane surfaces of glass, mica and platinum.},\ }\href@noop {} {\bibfield
  {journal} {\bibinfo  {journal} {J. Am. Chem. Soc}\ }\textbf {\bibinfo
  {volume} {40}},\ \bibinfo {pages} {1361} (\bibinfo {year}
  {1918})}\BibitemShut {NoStop}%
\bibitem [{\citenamefont {Liu}\ and\ \citenamefont
  {Shen}(2008)}]{liu2008langmuir}%
  \BibitemOpen
  \bibfield  {author} {\bibinfo {author} {\bibfnamefont {Y.}~\bibnamefont
  {Liu}}\ and\ \bibinfo {author} {\bibfnamefont {L.}~\bibnamefont {Shen}},\
  }\bibfield  {title} {\bibinfo {title} {From langmuir kinetics to first-and
  second-order rate equations for adsorption},\ }\href@noop {} {\bibfield
  {journal} {\bibinfo  {journal} {Langmuir}\ }\textbf {\bibinfo {volume}
  {24}},\ \bibinfo {pages} {11625} (\bibinfo {year} {2008})}\BibitemShut
  {NoStop}%
\bibitem [{\citenamefont {Hegemann}(2008)}]{hegemann2008algal}%
  \BibitemOpen
  \bibfield  {author} {\bibinfo {author} {\bibfnamefont {P.}~\bibnamefont
  {Hegemann}},\ }\bibfield  {title} {\bibinfo {title} {Algal sensory
  photoreceptors},\ }\href@noop {} {\bibfield  {journal} {\bibinfo  {journal}
  {Annu. Rev. Plant Biol.}\ }\textbf {\bibinfo {volume} {59}},\ \bibinfo
  {pages} {167} (\bibinfo {year} {2008})}\BibitemShut {NoStop}%
\bibitem [{\citenamefont {Harz}\ and\ \citenamefont
  {Hegemann}(1991)}]{harz1991rhodopsin}%
  \BibitemOpen
  \bibfield  {author} {\bibinfo {author} {\bibfnamefont {H.}~\bibnamefont
  {Harz}}\ and\ \bibinfo {author} {\bibfnamefont {P.}~\bibnamefont
  {Hegemann}},\ }\bibfield  {title} {\bibinfo {title} {Rhodopsin-regulated
  calcium currents in chlamydomonas},\ }\href@noop {} {\bibfield  {journal}
  {\bibinfo  {journal} {Nature}\ }\textbf {\bibinfo {volume} {351}},\ \bibinfo
  {pages} {489} (\bibinfo {year} {1991})}\BibitemShut {NoStop}%
\bibitem [{\citenamefont {Collingridge}\ \emph {et~al.}(2013)\citenamefont
  {Collingridge}, \citenamefont {Brownlee},\ and\ \citenamefont
  {Wheeler}}]{collingridge2013compartmentalized}%
  \BibitemOpen
  \bibfield  {author} {\bibinfo {author} {\bibfnamefont {P.}~\bibnamefont
  {Collingridge}}, \bibinfo {author} {\bibfnamefont {C.}~\bibnamefont
  {Brownlee}},\ and\ \bibinfo {author} {\bibfnamefont {G.~L.}\ \bibnamefont
  {Wheeler}},\ }\bibfield  {title} {\bibinfo {title} {Compartmentalized calcium
  signaling in cilia regulates intraflagellar transport},\ }\href@noop {}
  {\bibfield  {journal} {\bibinfo  {journal} {Current Biology}\ }\textbf
  {\bibinfo {volume} {23}},\ \bibinfo {pages} {2311} (\bibinfo {year}
  {2013})}\BibitemShut {NoStop}%
\bibitem [{\citenamefont {Fort}\ \emph {et~al.}(2021)\citenamefont {Fort},
  \citenamefont {Collingridge}, \citenamefont {Brownlee},\ and\ \citenamefont
  {Wheeler}}]{fort2021ca2+}%
  \BibitemOpen
  \bibfield  {author} {\bibinfo {author} {\bibfnamefont {C.}~\bibnamefont
  {Fort}}, \bibinfo {author} {\bibfnamefont {P.}~\bibnamefont {Collingridge}},
  \bibinfo {author} {\bibfnamefont {C.}~\bibnamefont {Brownlee}},\ and\
  \bibinfo {author} {\bibfnamefont {G.}~\bibnamefont {Wheeler}},\ }\bibfield
  {title} {\bibinfo {title} {Ca2+ elevations disrupt interactions between
  intraflagellar transport and the flagella membrane in chlamydomonas},\
  }\href@noop {} {\bibfield  {journal} {\bibinfo  {journal} {Journal of Cell
  Science}\ }\textbf {\bibinfo {volume} {134}},\ \bibinfo {pages} {jcs253492}
  (\bibinfo {year} {2021})}\BibitemShut {NoStop}%
\bibitem [{\citenamefont {Schneider}\ \emph {et~al.}(2015)\citenamefont
  {Schneider}, \citenamefont {Grimm}, \citenamefont {Hegemann} \emph
  {et~al.}}]{schneider2015biophysics}%
  \BibitemOpen
  \bibfield  {author} {\bibinfo {author} {\bibfnamefont {F.}~\bibnamefont
  {Schneider}}, \bibinfo {author} {\bibfnamefont {C.}~\bibnamefont {Grimm}},
  \bibinfo {author} {\bibfnamefont {P.}~\bibnamefont {Hegemann}}, \emph
  {et~al.},\ }\bibfield  {title} {\bibinfo {title} {Biophysics of
  channelrhodopsin},\ }\href@noop {} {\bibfield  {journal} {\bibinfo  {journal}
  {Annu. Rev. Biophys}\ }\textbf {\bibinfo {volume} {44}},\ \bibinfo {pages}
  {167} (\bibinfo {year} {2015})}\BibitemShut {NoStop}%
\bibitem [{\citenamefont {Sineshchekov}\ \emph {et~al.}(2002)\citenamefont
  {Sineshchekov}, \citenamefont {Jung},\ and\ \citenamefont
  {Spudich}}]{sineshchekov2002}%
  \BibitemOpen
  \bibfield  {author} {\bibinfo {author} {\bibfnamefont {O.~A.}\ \bibnamefont
  {Sineshchekov}}, \bibinfo {author} {\bibfnamefont {K.-H.}\ \bibnamefont
  {Jung}},\ and\ \bibinfo {author} {\bibfnamefont {J.~L.}\ \bibnamefont
  {Spudich}},\ }\bibfield  {title} {\bibinfo {title} {Two rhodopsins mediate
  phototaxis to low-and high-intensity light in chlamydomonas reinhardtii},\
  }\href@noop {} {\bibfield  {journal} {\bibinfo  {journal} {Proc. Natl. Acad.
  Sci. U. S. A.}\ }\textbf {\bibinfo {volume} {99}},\ \bibinfo {pages} {8689}
  (\bibinfo {year} {2002})}\BibitemShut {NoStop}%
\bibitem [{\citenamefont {Berthold}\ \emph {et~al.}(2008)\citenamefont
  {Berthold}, \citenamefont {Tsunoda}, \citenamefont {Ernst}, \citenamefont
  {Mages}, \citenamefont {Gradmann},\ and\ \citenamefont
  {Hegemann}}]{berthold2008channelrhodopsin}%
  \BibitemOpen
  \bibfield  {author} {\bibinfo {author} {\bibfnamefont {P.}~\bibnamefont
  {Berthold}}, \bibinfo {author} {\bibfnamefont {S.~P.}\ \bibnamefont
  {Tsunoda}}, \bibinfo {author} {\bibfnamefont {O.~P.}\ \bibnamefont {Ernst}},
  \bibinfo {author} {\bibfnamefont {W.}~\bibnamefont {Mages}}, \bibinfo
  {author} {\bibfnamefont {D.}~\bibnamefont {Gradmann}},\ and\ \bibinfo
  {author} {\bibfnamefont {P.}~\bibnamefont {Hegemann}},\ }\bibfield  {title}
  {\bibinfo {title} {Channelrhodopsin-1 initiates phototaxis and photophobic
  responses in chlamydomonas by immediate light-induced depolarization},\
  }\href@noop {} {\bibfield  {journal} {\bibinfo  {journal} {Plant Cell}\
  }\textbf {\bibinfo {volume} {20}},\ \bibinfo {pages} {1665} (\bibinfo {year}
  {2008})}\BibitemShut {NoStop}%
\end{thebibliography}
\end{document}